\documentclass[pre,twocolumn,aps,showpacs]{revtex4}
\newcommand{\bec}[1]{\mbox{\boldmath $ #1$}}
\usepackage{graphicx}
\begin{document}
\centerline{\bf PHYSICAL REVIEW E, v. 64, 056307 (2001)}
\bigskip
\bigskip
\title{Nonlinear turbulent magnetic diffusion and mean-field dynamo}
\author{Igor Rogachevskii}
\email{gary@menix.bgu.ac.il} \homepage{http://www.bgu.ac.il/~gary}
\author{Nathan Kleeorin}
\email{nat@menix.bgu.ac.il}
\affiliation{Department of Mechanical
Engineering, The Ben-Gurion
University of the Negev, \\
POB 653, Beer-Sheva 84105, Israel}
\date{\today}
\begin{abstract}
The nonlinear coefficients defining the mean electromotive force
({\em i.e.,} the nonlinear turbulent magnetic diffusion, the
nonlinear effective velocity, the nonlinear $ \kappa $-tensor,
{\em etc.}) are calculated for an anisotropic turbulence. A
particular case of an anisotropic background turbulence ({\em
i.e.,} the turbulence with zero mean magnetic field) with one
preferential direction is considered. It is shown that the
toroidal and poloidal magnetic fields have different nonlinear
turbulent magnetic diffusion coefficients. It is demonstrated that
even for a homogeneous turbulence there is a nonlinear effective
velocity which exhibits diamagnetic or paramagnetic properties
depending on anisotropy of turbulence and level of magnetic
fluctuations in the background turbulence. The diamagnetic
velocity results in the field is pushed out from the regions with
stronger mean magnetic field, while the paramagnetic velocity
causes the magnetic field tends to be concentrated in the regions
with stronger field. Analysis shows that an anisotropy of
turbulence strongly affects the nonlinear turbulent magnetic
diffusion, the nonlinear effective velocity and the nonlinear $
\alpha $-effect. Two types of nonlinearities (algebraic and
dynamic) are also discussed. The algebraic nonlinearity implies a
nonlinear dependence of the mean electromotive force on the mean
magnetic field. The dynamic nonlinearity is determined by a
differential equation for the magnetic part of the $ \alpha
$-effect. It is shown that for the $ \alpha \Omega $ axisymmetric
dynamo the algebraic nonlinearity alone (which includes the
nonlinear $ \alpha $-effect, the nonlinear turbulent magnetic
diffusion, the nonlinear effective velocity, {\em etc.}) cannot
saturate the dynamo generated mean magnetic field while the
combined effect of the algebraic and dynamic nonlinearities limits
the mean magnetic field growth.
\end{abstract}


\maketitle

\section{Introduction}

Generation of magnetic fields by turbulent flow of conducting fluid
is a fundamental problem which has a large number of applications in
solar physics and astrophysics, geophysics and planetary physics, {\em etc.}
In recent time the problem of nonlinear mean-field magnetic dynamo
is a subject of active discussions (see, {\em e.g.,}
\cite{KA92,GD94,BB96,K99,F99,MS99,RK2000,KMRS2000,UB2000,VC01}). It was
suggested in \cite{VC92} that the quenching of the nonlinear $ \alpha $-effect
is very strong and causes a very weak saturated mean magnetic
field. However, the later suggestion is in disagreement with observations of
galactic and solar magnetic fields (see, {\em e.g.,}
\cite{M78,P79,KR80,ZRS83,RSS88,S89}) and with numerical simulations (see, {\em e.g.,}
\cite{GR95,BN95,BR96}).

Saturation of the dynamo generated mean magnetic field is caused by
the nonlinear effects, {\em i.e.,} by the back reaction of the mean magnetic
field on the $ \alpha $-effect, turbulent magnetic diffusion,
differential rotation, {\em etc.} The evolution of the mean magnetic field
$ {\bf B} $ is determined by equation
\begin{eqnarray}
\partial {\bf B} / \partial t = \bec{\nabla} \times ({\bf V} \times {\bf B}
+ \bec{\cal E} - \eta \bec{\nabla} \times {\bf B}) \;,
\label{A12}
\end{eqnarray}
where $ {\bf V} $ is a mean velocity ({\em e.g.,} the
differential rotation), $ \eta $ is the magnetic diffusion due to the
electrical conductivity of fluid. The mean electromotive force
$ \bec{\cal E} = \langle {\bf u} \times {\bf b} \rangle $ in an
anisotropic turbulence is given by
\begin{eqnarray}
{\cal E}_{i} &=& \alpha_{ij} B_{j} + ({\bf V}^{\rm eff} {\bf
\times} {\bf B})_{i} - \eta_{ij} (\bec{\nabla} {\bf \times} {\bf
B})_{j}
\nonumber\\
&& - \kappa_{ijk} (\partial \hat B)_{jk} - [\bec{\delta} {\bf
\times} (\bec{\nabla} {\bf \times} {\bf B})]_{i} \; \label{A14}
\end{eqnarray}
(see \cite{R80,RKR00}), where $ (\partial \hat B)_{ij} = (1/2)
(\nabla_{i} B_{j} + \nabla_{j} B_{i}) ,$ $ {\bf u} $ and $ {\bf b} $ are
fluctuations of the velocity and magnetic field, respectively,
angular brackets denote averaging over an ensemble of turbulent fluctuations,
the tensors $ \alpha_{ij} $ and $ \eta_{ij} $ describe the
$ \alpha $-effect and turbulent magnetic diffusion, respectively,
$ {\bf V}^{\rm eff} $ is the effective diamagnetic (or paramagnetic)
velocity, $ \kappa_{ijk} $ and $ \bec{\delta} $ describe a nontrivial
behavior of the mean magnetic field in an anisotropic turbulence.
Nonlinearities in the mean-field dynamo imply dependencies of the
coefficients $ (\alpha_{ij} , \eta_{ij}, {\bf V}^{\rm eff}, $ {\em etc.})
defining the mean electromotive force on the mean magnetic field.
The $ \alpha $-effect and the differential rotation are the sources
of the generation of the mean magnetic field, while the
turbulent magnetic diffusion and the $ \kappa $-effect (which
is determined by the tensor $ \kappa_{ijk}) $ contribute to the
dissipation of the mean magnetic field.

In spite the nonlinear $ \alpha $-effect was under active study
(see, {\em e.g.,} \cite{F99,RK2000}), the nonlinear turbulent magnetic
diffusion, the nonlinear $ \kappa $-effect, the nonlinear diamagnetic
and paramagnetic effects, {\em etc.} are poorly understood.

In the present paper we derived equations for
the nonlinear turbulent magnetic diffusion,
the nonlinear effective velocity, the nonlinear $ \kappa $-effect, {\em etc.}
for an anisotropic turbulence. The obtained results for the nonlinear
mean electromotive force are specified for an
anisotropic background turbulence with one preferential direction.
The background turbulence is the turbulence with zero mean magnetic field.
We demonstrated that toroidal and poloidal magnetic fields have
different nonlinear turbulent magnetic diffusion coefficients.
It is shown that even for a homogeneous turbulence there is a
nonlinear effective velocity which can be a diamagnetic or paramagnetic
velocity depending on anisotropy of turbulence and level of magnetic
fluctuations in the background turbulence.

\section{The governing equations}

In order to derive equations for the nonlinear turbulent magnetic diffusion
and other nonlinear coefficients defining the mean electromotive force
we will use a mean field approach in which the magnetic, ${\bf H}$,
and velocity, ${\bf v}$, fields are divided into the mean and
fluctuating parts: $ {\bf H}={\bf B}+{\bf b}, \quad {\bf v}={\bf V}+{\bf u}$,
where the fluctuating parts have zero mean values,
$ {\bf V} = \langle {\bf v} \rangle = $ const, and  $ {\bf B} =
\langle {\bf H} \rangle .$ The momentum equation  and
the induction equation for the turbulent fields $ {\bf u} $ and $ {\bf
b} $ in a frame moving with a local velocity of the large-scale
flows $ {\bf V} $  are given by
\begin{eqnarray}
{\partial {\bf u} \over \partial t}  &=& - {\bec{\nabla} P' \over
\rho} - {1 \over \mu_{0} \rho} [ {\bf b} \times (\bec{\bf \nabla }
\times {\bf B }) + {\bf B} \times  ( \bec{\bf \nabla} \times {\bf
b}) ]
\nonumber \\
& & + {\bf T} + \nu \Delta {\bf u} + {\bf F} / \rho \;,
\label{Q4} \\
{\partial {\bf b} \over \partial t} &=& \bec{\bf \nabla } \times
({\bf u} \times {\bf B} - \eta  \bec{\bf \nabla } \times {\bf b})
+ {\bf G} \;, \label{Q5}
\end{eqnarray}
and $ \bec{\bf \nabla } {\bf \cdot} {\bf u} = 0 ,$ where $ P' $ are the fluctuations
of the hydrodynamic pressure, $ {\bf F} $ is a random external
stirring force, $ \nu $ is the kinematic viscosity, $ \eta $ is the magnetic
diffusion due to the electrical conductivity of fluid, $ \rho $ is
the density of fluid, $ \mu_{0} $ is the magnetic permeability of the
fluid, the nonlinear terms $ {\bf T} $ and $ {\bf G} $ are given by
$ {\bf T} = \langle ({\bf u} {\bf \cdot} \bec{\nabla}){\bf u} \rangle
- ({\bf u} {\bf \cdot} \bec{\nabla}){\bf u} +
[\langle {\bf b} \times ( \bec{\nabla} \times {\bf b}) \rangle
- {\bf b} \times (\bec{\nabla} \times {\bf b}) ] / (\mu_{0} \rho) ,$
and $ {\bf G} = \bec{\nabla} \times ({\bf u} \times {\bf b} - \langle
{\bf u} \times {\bf b} \rangle ) .$
We consider the case of large hydrodynamic $ ({\rm Re} = l_{0}
u_{0} / \nu \gg 1 ) $ and magnetic $ ({\rm Rm} = l_{0} u_{0} / \eta \gg 1 )$
Reynolds numbers, where $ u_{0} $ is the characteristic velocity in
the maximum scale $ l_{0} $ of turbulent motions.

\subsection{The procedure of the derivation of equation for the nonlinear
mean electromotive force}

The procedure of the derivation of equation for the nonlinear
mean electromotive force is as follows (for details, see Appendix A).

(a). By means of Eqs. (\ref{Q4}) and (\ref{Q5}) we derive equations for the
second moments:
\begin{eqnarray}
f_{ij}({\bf k, R}) &=& \int \langle u_i ({\bf k} + {\bf  K} / 2)
u_j(-{\bf k} + {\bf  K} / 2 ) \rangle
\nonumber \\
& & \times \exp(i {\bf K} {\bf \cdot} {\bf R}) \,d {\bf K} =
f_{ji}(-{\bf k, R}) \;,
\label{T10} \\
h_{ij}({\bf k, R}) &=& \int \langle b_i ({\bf k} + {\bf  K} / 2)
b_j(-{\bf k} + {\bf K} / 2) \rangle
\nonumber \\
& & \times \exp(i {\bf K} {\bf \cdot} {\bf R}) \,d {\bf K} /
\mu_{0} \rho = h_{ji}(-{\bf k, R})\;,
\label{T11} \\
g_{ij}({\bf k, R}) &=& \int \langle b_i ({\bf k} + {\bf  K} / 2)
u_j( -{\bf k} + {\bf  K}  / 2 ) \rangle
\nonumber \\
& & \times \exp(i {\bf K} {\bf \cdot} {\bf R}) \,d {\bf K} \; .
\label{T12}
\end{eqnarray}
where $ {\bf R} $ and $ {\bf K} $  correspond to the large scales, and
$ {\bf r} $ and $ {\bf k} $ to the  small ones, {\em i.e.,}
$ {\bf R} = ( {\bf x} +  {\bf y}) / 2  , \quad
{\bf r} =  {\bf x} - {\bf y}, \quad {\bf K} = {\bf k}_1 + {\bf k}_2,
\quad {\bf k} = ({\bf k}_1 - {\bf k}_2) / 2 .$

(b). We split all correlation functions ({\em i.e.,} $ f_{ij}, h_{ij},
g_{ij}) $ into two parts, {\em e.g.,} $ h_{ij} = h_{ij}^{(N)} + h_{ij}^{(S)} ,$
where the tensor $ h_{ij}^{(N)} = [h_{ij}({\bf k},{\bf R}) +
h_{ij}(-{\bf k},{\bf R})] / 2 $ describes the nonhelical part of the
tensor and $ h_{ij}^{(S)} = [h_{ij}({\bf k},{\bf R}) -
h_{ij}(-{\bf k},{\bf R})] / 2 $ determines the helical part of the
tensor. Such splitting is caused, {\em e.g.,} by different times
of evolution of the helical and nonhelical parts of the
magnetic tensor. In particular, the characteristic
time of evolution of the tensor $ h_{ij}^{(N)} $ is of the order $
\tau_{0} = l_{0} / u_{0} ,$ while the relaxation time of the tensor
$ h_{ij}^{(S)} $ is of the order of $ \tau_{0} {\rm Rm} $ (see, {\em e.g.,}
\cite{ZRS83,KR82,KRR94,KR99}).

(c). Equations for the second moments contain higher moments and
a problem of closing the equations for the higher moments arises.
Various approximate methods have been proposed for the solution of problems of
this type (see, {\em e.g.,} \cite{O70,MY75,Mc90}). The simplest procedure
is the $ \tau $-approximation, which is widely used in the theory
of kinetic equations. For magnetohydrodynamic turbulence this approximation
was used in \cite{PFL76} (see also \cite{RK2000,KRR90,KMR96}).
In the simplest variant, it allows us to express the third moments in
terms of the second moments:
\begin{eqnarray}
M_{ij} - M_{ij}^{(0)} &=& - (f_{ij} - f_{ij}^{(0)}) / \tau (k) \;,
\label{A1} \\
R_{ij}^{(N)} &=& - (h_{ij}^{(N)} - h_{ij}^{(0N)}) / \tau (k) \;,
\label{A2} \\
C_{ij} &=& - g_{ij} / \tau (k) \;,
\label{A3}
\end{eqnarray}
where $ M_{ij} ,$ $ R_{ij} ,$ $ C_{ij} $ are the third moments in
equations for $ f_{ij}, h_{ij} $ and $ g_{ij} ,$ respectively
(see Eqs. (\ref{R3})-(\ref{Q3}) in Appendix A). The superscript
$ {(0)} $ corresponds to the background magnetohydrodynamic turbulence
(it is a turbulence with zero mean magnetic field, $ {\bf B} = 0),$
$ h_{ij}^{(0N)} $ is the nonhelical part of the tensor of magnetic
fluctuations of the background turbulence,
and $ \tau (k) $ is the characteristic relaxation time of the statistical
moments. We applied the $ \tau $-approximation  only for the
nonhelical part $ h_{ij}^{(N)} $ of the tensor of magnetic fluctuations
because the corresponding helical part $ h_{ij}^{(S)} $ is determined
by an evolutionary equation (see, {\rm e.g.,}
\cite{ZRS83,KR82,VK83,KRR94,GD94,KR99,KMRS2000}
and Section III-C). We took into account
here magnetic fluctuations which can be generated by a
stretch-twist-fold mechanism when a mean magnetic field is zero
(see, {\em e.g.,} \cite{ZRS90,RK97}). This implies that $ h_{ij}^{(0)}
\not= 0 .$ In inertia range of background turbulence $ R_{ij}({\bf B}=0) = 0 $
and $ C_{ij}({\bf B}=0) = 0 .$ We also took into account that
the cross-helicity tensor $ g_{ij} $ for $ {\bf B}=0 $ is zero,
{\em i.e.,} $ g_{ij}({\bf B}=0) = 0 .$

The $ \tau $-approximation  is in general similar to Eddy Damped
Quasi Normal Markovian (EDQNM) approximation. However some
principle difference exists between these two approaches (see
\cite{O70,Mc90}). The EDQNM closures do not relax to equilibrium,
and this procedure does not describe properly the motions in the
equilibrium state in contrast to the $ \tau $-approximation. Within
the EDQNM theory, there is no dynamically determined relaxation time,
and no slightly perturbed steady state can be approached \cite{O70}.
In the $ \tau $-approximation, the relaxation time for small
departures from equilibrium is determined by the random motions in
the  equilibrium state, but not by the departure from equilibrium
\cite{O70}. We use the $ \tau $-approximation, but not the EDQNM
approximation because we consider a case with $ l_{0} \vert
\bec{\nabla} B^2 \vert / \mu_{0}
\ll \langle  \rho u^2 \rangle .$ As follows from the analysis
by \cite{O70} the $ \tau $-approximation describes the relaxation to
equilibrium state (the background turbulence) much more accurately than
the EDQNM approach.

In this study we consider an intermediate nonlinearity which implies that the
mean magnetic field is not enough strong in order to affect the correlation
time of turbulent velocity field. The theory for a very strong
mean magnetic field can be corrected after taking into account a
dependence of the correlation time of the turbulent velocity field on
the mean magnetic field.

(d). We assume that the characteristic time of variation of the mean magnetic
field $ {\bf B} $ is substantially larger than the correlation time
$ \tau(k) $ for all turbulence scales. This allows us to get a stationary
solution for the equations for the second moments
$ f_{ij}, h_{ij} $ and $ g_{ij} .$
Using these equations [see Eqs. (\ref{Q39})-(\ref{Q70}) in Appendix A]
we calculate the electromotive force
$ {\cal E}_{i}({\bf r}=0) = \int {\cal E}_{i}({\bf k}) \,d {\bf k} ,$
where $ {\cal E}_{i}({\bf k}) = (1/2) \varepsilon_{imn}
(g_{nm}^{(N)}({\bf k, R}) - g_{mn}^{(N)}(-{\bf k},{\bf R})) .$
The result is given by
\begin{eqnarray}
{\cal E}_{i}({\bf r}=0) = a_{ij} B_{j} + b_{ijk} B_{j,k} \;,
\label{P20}
\end{eqnarray}
where $ B_{i,j} = \partial B_{i} / \partial R_{j} ,$
\begin{eqnarray}
a_{ij} &=& i \int \tau (1 + \psi)^{-1} \varepsilon_{imn} k_{j} (f_{nm}^{(0S)}
- h_{nm}^{(S)}) \,d {\bf k} \;,
\label{P21} \\
b_{ijk} &=& \int \tau (1 + \psi)^{-1} [\varepsilon_{ijn} (f_{kn}^{(0N)} +
h_{kn}^{(0N)})
\nonumber \\
& & - 2 \varepsilon_{imn} k_{mj} h_{nk}^{(N)} ] \,d {\bf k} \;
\label{P22}
\end{eqnarray}
(for details, see Appendix A), $ k_{ij} = k_{i} k_{j} / k^{2} ,$
$ h_{nm}^{(N)} = h_{nm}^{(0N)} + \psi
(1 + 2 \psi)^{-1} (f_{nm}^{(0N)} - h_{nm}^{(0N)}) ,$ $ \varepsilon_{ijk} $
is the Levi-Civita tensor, and $ \psi = [(\bec{\beta} {\bf \cdot} {\bf k}) u_{0}
\tau / 2]^{2} ,$ $ \beta_{i} = 4 B_{i} / (u_{0} \sqrt{2 \mu_{0} \rho}) ,$
$ f_{ij}^{(0N)} $ and $ f_{ij}^{(0S)} $ describe the nonhelical and helical
tensors of the background turbulence.

(e). Following to \cite{R80} we use an identity
$ B_{j,k} = (\partial \hat B)_{jk} - \varepsilon_{jkl}
(\bec{\bf \nabla} {\bf \times} {\bf B})_{l} / 2 $
which allows us to rewrite Eq. (\ref{P20}) for the electromotive force
in the form
\begin{eqnarray}
{\cal E}_{i} = {\alpha}_{ij} B_{j} + ({\bf U} {\bf \times} {\bf B})_{i}
- {\eta}_{ij} (\bec{\nabla} {\bf \times} {\bf B})_{j}
- {\kappa}_{ijk} ({\partial \hat B})_{jk} \;,
\label{PPP23}
\end{eqnarray}
where
\begin{eqnarray}
\alpha_{ij}({\bf B}) &=& (a_{ij} + a_{ji}) / 2 \;,
\quad  U_{k}({\bf B}) = \varepsilon_{kji} a_{ij} / 2 \;,
\label{DDD41} \\
\eta_{ij} &=& (\varepsilon_{ikp} b_{jkp} + \varepsilon_{jkp} b_{ikp}) / 4 \;,
\nonumber \\
\kappa_{ijk}({\bf B}) &=& - (b_{ijk} + b_{ikj}) / 2 \; .
\label{DDD40}
\end{eqnarray}

\subsection{The model for the background turbulence}

For the integration in $ {\bf k} $-space in Eqs. (\ref{P21}) and (\ref{P22})
we have to specify a model for the background turbulence ({\em i.e.,}
turbulence with zero mean magnetic field). We assume that the background
turbulence is anisotropic and incompressible. The second moments
for turbulent velocity and magnetic fields of the background
turbulence are given by
\begin{eqnarray}
\tau c_{ij}({\bf k}) &=& (5/4) \{ P_{ij}(k) [(2/5) \tilde
\eta_{T}^{(a)}({\bf k}) - \mu_{mn}^{(a)}({\bf k}) k_{nm}]
\nonumber\\
& & + 2 [\delta_{ij} \mu_{mn}^{(a)}({\bf k}) k_{nm} +
\mu_{ij}^{(a)}({\bf k}) - \mu_{im}^{(a)}({\bf k}) k_{mj}
\nonumber\\
& & - k_{im} \mu_{mj}^{(a)}({\bf k})] \} \; \label{Q19}
\end{eqnarray}
(see \cite{RK2000}), where $ c_{ij} = f_{ij}^{(0N)} $ when $ a = v ,$ and
$ c_{ij} = h_{ij}^{(0N)} $ when $ a = h ,$
and $ \tilde \eta_{T}^{(v)}({\bf k}) = \tau f_{pp}^{(0N)}({\bf k}) ,$
$ \quad \tilde \eta_{T}^{(h)}({\bf k}) = \tau h_{pp}^{(0N)}({\bf k}) ,$
$ P_{ij}(k) =  \delta _{ij} - k_{ij} ,$ $ \delta_{mn} $ is the Kronecker
tensor. The anisotropic part of this
tensor $ \mu_{mn}^{(a)}({\bf k}) $ has the properties:
$ \mu_{mn}^{(a)}({\bf k}) = \mu_{nm}^{(a)}({\bf k}) $ and
$ \mu_{pp}^{(a)}({\bf k}) = 0 .$ Inhomogeneity of the background turbulence
is assumed to be weak, {\em i.e.,} in Eq. (\ref{Q19}) we dropped terms
$ \sim O[\bec{\nabla}(\eta_{T}^{(a)};\mu_{ij}^{(a)})] ,$
where $ \eta_{T}^{(v)} = \tau_{0} u_{0}^{2} / 3 ,$ $ \eta_{T}^{(h)} =
\tau_{0} b_{0}^{2} / 3 \mu_{0} \rho $ and $ b_{0} $ is the
characteristic value of the magnetic fluctuations in the background
turbulence. To integrate over $ k $ in Eqs. (\ref{P21}) and (\ref{P22})
we use the Kolmogorov spectrum of the background turbulence,
{\em i.e.,} $ \tau f_{pp}^{(0N)}({\bf k}) = \eta_{T}^{(v)} \varphi(k) ,$
$ \tau h_{pp}^{(0N)}({\bf k}) = \eta_{T}^{(h)} \varphi(k) $ and
$ \mu_{mn}^{(a)}({\bf k}) = \mu_{mn}^{(a)}({\bf R}) \varphi(k) / 3 ,$
where $ \varphi(k) = (\pi k^{2} k_{0})^{-1} (k / k_{0})^{-7/3} ,$
$ \tau(k) = 2 \tau_{0} (k / k_{0})^{-2/3} ,$ $ \quad  k_{0} = l_{0}^{-1} .$
We take into account that the inertial range
of the turbulence exists in the scales: $ l_{d} \leq r \leq l_{0} .$ Here
the maximum scale of the turbulence $ l_{0} \ll L_{B} ,$ and $ l_{d} = l_{0} /
{\rm Re}^{3/4} $ is the viscous scale of turbulence, and
$ L_{B} $ is the characteristic scale of variations of the nonuniform
mean magnetic field.

In the next section we present results for the nonlinear coefficients
defining the mean electromotive force.

\section{Nonlinear coefficients defining the mean electromotive force}

The procedure described in Section II (see also, for details
Appendix A) allows us to calculate the nonlinear turbulent magnetic diffusion
tensor, the nonlinear $ \bec{\kappa} $-tensor, the nonlinear $
\bec{\alpha} $-tensor and the nonlinear effective drift velocity.

\subsection{Nonlinear turbulent magnetic diffusion tensor
and nonlinear $ \kappa $-tensor}

The general form of the turbulent magnetic diffusion tensor
$ \eta_{ij}({\bf B}) $ contains all possible tensors: $
\delta_{ij} ,$ $ \mu_{ij}^{(a)} $ $ \beta_{ij} $ and their symmetric
combination $ \bar \mu_{ij}^{(a)} = \mu_{in}^{(a)} \beta_{nj} + \beta_{in}
\mu_{nj}^{(a)} $  [see Eq. (\ref{D20}) in Appendix A], where
$ \beta_{ij} = \beta_{i} \beta_{j} / \beta^{2} ,$
$ \beta_{i} = 4 B_{i} / (u_{0} \sqrt{2 \mu_{0} \rho}) .$ For an isotropic
background turbulence (when $ \mu_{ij}^{(a)} = 0) $ the turbulent magnetic
diffusion tensor $ \eta_{ij}({\bf B}) $ is given by
\begin{eqnarray}
\eta_{ij}({\bf B}) &=& \delta_{ij} \{ A_{1}(\sqrt{2}\beta)
\eta_{T}^{(v)} + [A_{1}(\beta) - A_{1}(\sqrt{2}\beta)]
\eta_{T}^{(h)} \}
\nonumber\\
& & + {1 \over 2} \beta_{ij} A_{2}(\beta) (\eta_{T}^{(v)} +
\eta_{T}^{(h)}) \;, \label{PD20}
\end{eqnarray}
where the functions $ A_{k}(\beta) $ are defined in Appendix B. For
$ \beta \ll 1 $ Eq. (\ref{PD20}) reads
\begin{eqnarray}
\eta_{ij}({\bf B}) &=& \delta_{ij} [\eta_{T}^{(v)} - (2
\beta^{2}/5) (2 \eta_{T}^{(v)} - \eta_{T}^{(h)})]
\nonumber\\
& & - (2/5) \beta_{i} \beta_{j} (\eta_{T}^{(v)} + \eta_{T}^{(h)})
\;, \label{PD21}
\end{eqnarray}
and for $ \beta \gg 1 $ it is given by
\begin{eqnarray}
\eta_{ij}({\bf B}) &=& (3 \pi /10 \beta) \{ \sqrt{2} \delta_{ij}
[\eta_{T}^{(v)} + \eta_{T}^{(h)}(\sqrt{2} - 1)]
\nonumber\\
& & - \beta_{ij} (\eta_{T}^{(v)} + \eta_{T}^{(h)}) \} \; .
\label{PD22}
\end{eqnarray}
The mean magnetic field causes an anisotropy of the turbulent magnetic diffusion
tensor which is determined by the tensor $ \beta_{ij} .$ Magnetic
fluctuations of the background turbulence contribute to the turbulent magnetic
diffusion tensor $ \eta_{ij}({\bf B}) $ in the nonlinear case.
It follows from Eq. (\ref{PD22}) that for $ \beta \gg 1 $ the tensor
$ \eta_{ij} \propto 1 / \beta .$

The $ \kappa $-tensor describes a nontrivial
behavior of the mean magnetic field in an anisotropic turbulence.
For an isotropic background turbulence the $ \kappa $-tensor vanishes
in spite of an anisotropy caused by the mean magnetic field.
For an anisotropic background turbulence a general form of the
$ \kappa $-tensor is given by Eq. (\ref{D24}) in Appendix A.
For $ \beta \ll 1 $ this tensor is given by
\begin{eqnarray}
\kappa_{ijk} &=& - (1/6) (3 \hat L_{ijk}^{(v)} + \hat
L_{ijk}^{(h)}) + (1/7) \beta^{2} (5 \hat L_{ijk}^{(v)} + \hat
L_{ijk}^{(h)}
\nonumber\\
& & - 4 \hat N_{ijk}^{(v)} + 2 \hat N_{ijk}^{(h)}) \;,
\label{PD25}
\end{eqnarray}
and for $ \beta \gg 1 $ it reads
\begin{eqnarray}
\kappa_{ijk} &=& - (\pi / 16 \beta) (\sqrt{2} - 1) [\hat
L_{ijk}^{(v)} + \hat L_{ijk}^{(h)}
\nonumber\\
& & + 3(\hat N_{ijk}^{(v)} + \hat N_{ijk}^{(h)})] \;, \label{PD30}
\end{eqnarray}
where
$ \hat L_{ijk}^{(a)} = \varepsilon_{ijn} \mu_{nk}^{(a)} +
\varepsilon_{ikn} \mu_{nj}^{(a)} $ and $ \quad \hat N_{ijk}^{(a)} =
\mu_{np}^{(a)} (\varepsilon_{ijn} \beta_{pk} +
\varepsilon_{ikn} \beta_{pj}) .$ Note that for $ \beta \gg 1 $ the tensor
$ \kappa_{ijk} \propto 1 / \beta .$ The $ \kappa $-tensor contributes to
the turbulent magnetic diffusion of the toroidal and poloidal mean magnetic
fields (see Section V).

\subsection{The hydrodynamic part of the nonlinear $ \alpha $-tensor}

Using Eqs. (\ref{P21}) and (\ref{DDD41}) we get
\begin{eqnarray}
\alpha_{ij}^{(v)}({\bf B},{\bf R}) = \int {\alpha_{ij}^{(v)}(0,{\bf k},{\bf R})
\over 1 + \psi({\bf B},{\bf k})} \,d{\bf k}  \;,
\label{C1}
\end{eqnarray}
where hereafter $ \alpha_{ij}^{(v)}(0,{\bf k},{\bf R}) \equiv
\alpha_{ij}^{(v)}({\bf B}=0,{\bf k},{\bf R}) .$ Analysis in \cite{RK2000,RKR00}
shows that a form of the tensor $ \alpha_{ij}^{(v)}(0,{\bf k},{\bf R}) $ in an
anisotropic turbulence can be constructed using the tensors $ k_{ij} ,$
$ k_{ijmn} $ and $ \nu_{ij} ,$ where $ k_{ijmn} = k_{ij} k_{mn} $ and
$ \nu_{ij} $ is the anisotropic part of the hydrodynamic contribution
of the $ \alpha $-tensor. Thus we use the following model
for the tensor $ \alpha_{ij}^{(v)}(0,{\bf k},{\bf R}) $
\begin{eqnarray}
\alpha_{ij}^{(v)}(0,{\bf k},{\bf R}) = \{ 2 \alpha_{0}^{(v)}({\bf
R}) k_{ij} + 5 \epsilon k_{ijmn} \nu_{mn}({\bf R})
\nonumber\\
+ (1 - \epsilon)[\nu_{ip}({\bf R}) k_{pj} + \nu_{jp}({\bf R})
k_{pi}] \} \varphi(k) / 2 \;, \label{C2}
\end{eqnarray}
where the parameter $ \epsilon $ describes an anisotropy of the helical component
of turbulence and it changes in the interval: $ 0 \leq \epsilon \leq 1 .$ Here
$ \alpha_{ij}^{(v)}(0,{\bf R}) = \int \alpha_{ij}^{(v)}(0,{\bf k},{\bf R})
\,d{\bf k} = \alpha_{0}^{(v)}({\bf R}) \delta_{ij} + \nu_{ij}({\bf R}) ,$
and $ \alpha_{0}^{(v)}({\bf R}) = (1/3) \alpha_{pp}^{(v)}(0,{\bf R}) ,$
the anisotropic part $ \nu_{ij}({\bf R}) $ of the hydrodynamic
contribution of the $ \alpha $-tensor has the properties:
$ \nu_{ij} = \nu_{ji} $ and $ \nu_{pp} = 0 .$ Substituting Eq. (\ref{C2}) into
Eq. (\ref{C1}), and using identities (\ref{C22}) and (\ref{C24})
we obtain the nonlinear dependence of the hydrodynamic part
of the $ \alpha $-effect on mean magnetic field:
\begin{eqnarray}
\alpha_{ij}^{(v)}({\bf B}) &=& (1/2) \{\delta_{ij} [2
(A_{1}(\beta) + A_{2}(\beta)) \alpha_{0}^{(v)}
\nonumber\\
&& + (1 - \epsilon) A_{2}(\beta) \nu_{\beta} + 5 \epsilon
(C_{2}(\beta) + 3 C_{3}(\beta)) \nu_{\beta}]
\nonumber\\
&& + \nu_{ij} [(1 - \epsilon) (2 A_{1}(\beta) + A_{2}(\beta))
\nonumber\\
&& + 10 \epsilon (C_{1}(\beta) + C_{2}(\beta))] \} \;, \label{C4}
\end{eqnarray}
where $ \nu_{\beta}({\bf R}) = \nu_{mn}({\bf R}) \beta_{nm} $
and the functions $ C_{k}(\beta) $ are defined in Appendix B.
For $ \epsilon = 0 $ Eq. (\ref{C4}) coincides with that
derived in \cite{RK2000}. The asymptotic formulas for
$ \alpha_{ij}^{(v)} $ for $ \beta \ll 1 $ and
$ \beta \gg 1 $ are given by Eqs. (\ref{C5}) and (\ref{C6})
in Appendix A.

\subsection{The mean electromotive force and the nonlinear
magnetic $ \alpha $-tensor}

Using Eqs. (\ref{DD15}), (\ref{D18}), (\ref{P23}) we calculate
the electromotive force $ \bec{\cal E} $
\begin{eqnarray}
{\cal E}_{i} &=& {\alpha}_{ij} B_{j} + ({\bf V}^{\rm eff} {\bf
\times} {\bf B})_{i} - {\eta}_{ij} (\bec{\nabla} {\bf \times} {\bf
B})_{j}
\nonumber\\
&& - {\kappa}_{ijk} ({\partial \hat B})_{jk} \;, \label{D27}
\end{eqnarray}
where the nonlinear effective drift velocity $ {\bf V}^{\rm eff} = {\bf U}
+ {\bf V}^{(N)} ,$ and the velocity $ U_{i}({\bf B}) = - (1/2)
\varepsilon_{imn} a_{mn} = - (1/2) {\nabla}_{p} \Lambda^{(M)}_{pi}
(\sqrt{2} \beta) $ (see \cite{RK2000}), the velocity $ {\bf V}^{(N)} $
is given by Eq. (\ref{D19}),
the tensor of turbulent magnetic diffusion $ \eta_{ij} $ is given by Eq.
(\ref{D20}), the tensor $ \kappa_{ijk} $ is determined by Eq. (\ref{D24}),
the tensor $ \Lambda^{(M)}_{ij} $ is defined in Eqs. (\ref{DD2}) and
(\ref{P43}). In the kinematic dynamo the effective drift velocity (turbulent
diamagnetic velocity) is caused by an inhomogeneity of turbulence.
The effective drift velocity $ {\bf U}({\bf B}) $ is determined by the tensor
$ a_{ij} $ and is due to an induced inhomogeneity of turbulence
caused by the nonuniform mean magnetic field. This implies that
the nonuniform mean magnetic field modifies turbulent velocity field
and creates the inhomogeneity of turbulence. The effective velocity
$ {\bf V}^{(N)}({\bf B}) $ is determined by tensor $ b_{ijk} $  and is caused by
the nonuniform mean magnetic field.

The $ \alpha $-tensor is determined by the hydrodynamic and
magnetic contributions, {\em i.e.,} $ \alpha_{ij}({\bf B}) =
\alpha_{ij}^{(v)}({\bf B}) + \alpha_{ij}^{(h)}({\bf B}) $ with
\begin{eqnarray}
\alpha_{ij}^{(h)}({\bf B}) = \alpha_{0}^{(h)}({\bf B}) \Phi(\beta)
\delta_{ij}
\label{D28}
\end{eqnarray}
(see \cite{RK2000}), where the tensor $ \alpha_{ij}^{(v)}({\bf B}) $
is determined by Eq. (\ref{C4}), the function $ \Phi(\beta) = (3 / \beta^{2})
[1 - \arctan (\beta) / \beta] ,$ and the magnetic part
$ \alpha^{(h)}_{0}({\bf B}) $ of the $ \alpha $-effect is determined by
the dynamic equation
\begin{eqnarray}
{\partial \alpha^{(h)}_{0} \over \partial t} + {\alpha^{(h)}_{0} \over T} +
\bec{\nabla} {\bf \cdot} ({\bf W} \alpha^{(h)}_{0} + {\bf F}_{\rm flux})
\nonumber\\
= - {4 \over 9 \eta_{T} \mu_{0} \rho} \bec{\cal E}({\bf B}) {\bf
\cdot} {\bf B} \; \label{A4}
\end{eqnarray}
(see \cite{KMRS2000,KR99,KR82,VK83,KRR94,RK99}), where $ W_{i} =
c_{ij} V_{j} $ is the velocity which depends on the mean fluid
velocity $ {\bf V} $ (for an isotropic turbulence the tensor
$ c_{ij} = \delta_{ij} $ and for an anisotropic turbulence with one
preferential direction, say in the direction $ {\bf e} ,$ the tensor
$ c_{ij} = (23/30) \delta_{ij} + (7/10) e_{i} e_{j} ,$ see
\cite{KR99}); the flux
\begin{eqnarray}
{\bf F}_{\rm flux} \propto \tau \alpha^{(v)}({\bf B}) {\bec{\nabla} \rho
\over \rho} \biggl({\eta_{T}^{(v)}({\bf B}) \, B^{2} \over \eta_{T}^{(v)}({\bf
B}=0) \, \mu_{0} \rho} \biggr) \;
\end{eqnarray}
is related with the flux of the magnetic helicity and is independent
of the mean fluid velocity $ {\bf V} $ \cite{KMRS2000} (see also \cite{VC01}),
and $ T \sim \tau_{0} {\rm Rm} $ is the characteristic time of relaxation
of magnetic helicity.
The asymptotic formulas for $ \alpha_{ij}^{(h)} $ for $ \beta \ll 1 $ and
$ \beta \gg 1 $ are given by Eqs. (\ref{D30}) and (\ref{D32})
in Appendix A.

\section{Anisotropic background turbulence with one preferential direction}

Now we consider an anisotropic background turbulence with one preferential
direction, say along an unit vector $ {\bf e} .$ Thus the tensor $
\eta_{ij}^{(v)}({\bf B}=0) = \langle \tau v_{i} v_{j} \rangle $ is given by
$ \eta_{ij}^{(v)}({\bf B}=0) = \eta_{T}^{(v)} \delta_{ij} + \mu_{ij}^{(v)} =
\eta_{0}^{(v)} \delta_{ij} + \varepsilon_{\mu}^{(v)} e_{ij} ,$
where the trace $ \eta_{pp}^{(v)}({\bf B}=0) $ in this equation yields
$ \eta_{0}^{(v)} = \eta_{T}^{(v)} - (1/3) \varepsilon_{\mu}^{(v)} $ and
$ e_{ij} = e_{i} e_{j} .$ Therefore, the anisotropic part $ \mu_{ij}^{(v)} $
of the tensor $ \eta_{ij}^{(v)}({\bf B}=0) $ is given by
$ \mu_{ij}^{(v)} = \varepsilon_{\mu}^{(v)} (e_{ij} - (1/3) \delta_{ij}) ,$
and $ \bar \mu_{ij}^{(v)} \equiv \mu_{in}^{(v)} \beta_{nj} + \beta_{in}
\mu_{nj}^{(v)} = \varepsilon_{\mu}^{(v)} [(e_{i} \beta_{j} +
e_{j} \beta_{i}) ({\bf e} {\bf \cdot} \bec{\hat \beta}) - (2/3) \beta_{ij}] ,$
where $ \varepsilon_{\mu}^{(v)} $ is a degree of of an anisotropy of the
turbulence, and $ \mu_{\beta}^{(v)} \equiv (1/2) \bar \mu_{pp}^{(v)}
= \varepsilon_{\mu}^{(v)} [({\bf e} {\bf \cdot}
\bec{\hat \beta})^{2} - 1/3] .$ It follows from these equations that
$ \eta_{ij}^{(v)}({\bf B}=0) = \delta_{ij} [\eta_{T}^{(v)} -
(1/3) \varepsilon_{\mu}^{(v)}] + e_{ij} \varepsilon_{\mu}^{(v)} .$
Now we take into account that the components $ \eta_{xx}^{(v)}({\bf B}=0) ,$
$ \eta_{yy}^{(v)}({\bf B}=0) $ and $ \eta_{zz}^{(v)}({\bf B}=0) $
are positive. This yields
$ -3/2 \leq \varepsilon_{\mu}^{(v)} / \eta_{T}^{(v)} \leq 3 .$
The equations for the corresponding magnetic tensors are obtained from
these equations after the change $ v \to h .$ For the magnetic fluctuations
we also obtain that $ -3/2 \leq \varepsilon_{\mu}^{(h)}
/ \eta_{T}^{(h)} \leq 3 .$

For galaxies, {\em e.g.,} the preferential direction $ {\bf e} $ is along
rotation (which is parallel to the effective gravity field).
For the axisymmetric $ \alpha \Omega $--dynamo and large magnetic Reynolds
numbers the toroidal magnetic field is much larger than the poloidal
field. Therefore, the value $ {\bf e} {\bf \cdot} \bec{\hat \beta} $ is very small
and can be neglected because $ \bec{\hat \beta} $ is approximately
directed along the toroidal magnetic field.

Thus, the nonlinear coefficients defining the mean electromotive force
in a turbulence with one preferential direction are given by
\begin{eqnarray}
\eta_{ij}({\bf B}) = M_{\eta} \delta_{ij} + M_{e} e_{ij} +
M_{\beta} \beta_{ij} \;,
\label{D47} \\
{\bf V}^{\rm eff}({\bf B})  = B^{-2} [M_{V}^{(1)} \bec{\nabla}
B^{2} + M_{V}^{(2)} {\bf e} ({\bf e} {\bf \cdot} \bec{\nabla})
B^{2}] \;,
\label{D48} \\
\kappa_{ijk}({\bf B}) (\partial \hat B)_{jk} = M_{\kappa} [{\bf e}
{\bf \times}({\bf e} {\bf \cdot} \bec{\nabla}) {\bf B}]_{i} \;
\label{D49}
\end{eqnarray}
(see Appendix C), where we assumed that $ {\bf e} {\bf \cdot} \bec{\hat \beta} = 0 ,$
the functions $ M_{\eta} ,$ $ M_{e} ,$ $ M_{\beta} ,$
$ M_{\kappa} ,$ $ M_{V}^{(1)} $ and $ M_{V}^{(2)} $ are given by Eqs.
(\ref{D50})-(\ref{D54}) in Appendix C. The tensor $ \eta_{ij}({\bf B}) $
contains three tensors $ \delta_{ij} ,$ $ e_{ij} $ and $ \beta_{ij} $
since here there two preferred directions, along the vectors $ {\bf e} $
and $ {\bf B} .$

Now we consider the hydrodynamic part of the $ \alpha $-effect
for an anisotropic background turbulence with one preferential
direction. The tensor $ \alpha_{ij}^{(v)}({\bf B}=0) $ in this case can be
rewritten in the form
$ \alpha_{ij}^{(v)}({\bf B}=0) = \alpha_{0}^{(v)} \delta_{ij} +
\nu_{ij} = [\alpha_{0}^{(v)} - (1/3) \varepsilon_{\alpha}] \delta_{ij}
+ \varepsilon_{\alpha} e_{ij} ,$
where $ \varepsilon_{\alpha} $ is a degree of an anisotropy of the
$ \alpha $-tensor. Thus, the anisotropic part $ \nu_{ij} $ is given by
$ \nu_{ij} = \varepsilon_{\alpha} (e_{ij} - (1/3) \delta_{ij}) .$
The electromotive force contains the tensor $ \alpha_{ij} $
in the form $ \alpha_{ij} B_{j} .$ Thus,
$ \nu_{ij} \hat \beta_{j} = - (1/3) \varepsilon_{\alpha}
[\hat \beta_{i} - 3 ({\bf e} {\bf \cdot} \bec{\hat \beta}) e_{i}] $ and
$ \nu_{\beta} = - (1/3) \varepsilon_{\alpha}
[1 - 3 ({\bf e} {\bf \cdot} \bec{\hat \beta})^{2}] .$
Using Eq. (\ref{C4}) we obtain
the hydrodynamic part of the $ \alpha $-tensor in
an anisotropic background turbulence with one preferential
direction
\begin{eqnarray}
\alpha_{ij}^{(v)}({\bf B}) &=& \delta_{ij} \{
[A_{1}(\beta) + A_{2}(\beta)] [\alpha_{0}^{(v)} - (1/3) \varepsilon_{\alpha}
(1 - \epsilon)]
\nonumber \\
&& - (5/6) \varepsilon_{\alpha} \epsilon [2 C_{1}(\beta) + 3 C_{2}(\beta)
+ 3 C_{3}(\beta)] \}
\nonumber\\
&& \equiv \Phi_{\alpha}(\beta) \delta_{ij} \;, \label{C8}
\end{eqnarray}
where we assumed that $ {\bf e} {\bf \cdot} \bec{\hat \beta} = 0 .$
For $ \epsilon \not=0 $ the tensor $ \alpha_{ij}^{(v)}({\bf B}) $ can change
its sign at some value $ B_{\ast} $ of the mean magnetic field [see
Eqs. (\ref{C9}) and (\ref{C10}) in Appendix C]. Thus the point
$ B = B_{\ast} $ can determine a steady state configuration of the
mean magnetic field for $ \epsilon \not=0 .$

\section{Applications: Mean-field equations for the thin-disk axisymmetric
$ \alpha \Omega $--dynamo}

Here we apply the obtained results for the nonlinear mean electromotive
force to the analysis of the thin-disk axisymmetric $ \alpha \Omega $--dynamo.
Using Eqs. (\ref{D47})-(\ref{D49}) we derive the mean-field equations
for the thin-disk axisymmetric $ \alpha \Omega $--dynamo:
\begin{eqnarray}
{\partial B \over \partial t} &=& {\partial \over \partial z} \biggl(
\eta_{B} {\partial B \over \partial z} \biggr) + \tilde G {\partial A \over
\partial z} \;,
\label{D55} \\
{\partial A \over \partial t} &=& \eta_{A} {\partial^{2} A \over
\partial z^{2}} - V_{A} {\partial A \over \partial z} + \alpha B \;,
\label{D56}
\end{eqnarray}
where $ r ,$ $ \varphi $ and $ z $ are cylindrical coordinates,
$ {\bf B} = B {\bf e}_{\varphi} + \bec{\nabla} {\bf \times}
(A {\bf e}_{\varphi}) ,$ $ \tilde G = - r (\partial \Omega / \partial r) ,$ and
\begin{eqnarray}
\eta_{A}({\bf B}) &=& M_{\eta} + M_{\kappa} + M_{\beta} \;,
\nonumber\\
\eta_{B}({\bf B}) &=& M_{\eta} + M_{\kappa} - 2 M_{V} \;,
\label{D58} \\
V_{A}({\bf B}) &=& (\eta_{A} - \eta_{B}) (\ln \vert B \vert)' \;,
\nonumber\\
\alpha({\bf B}) &=& \Phi_{\alpha}({\bf B}) \alpha_{0}^{(v)} +
\Phi({\bf B}) \alpha_{0}^{(h)}({\bf B}) \;, \label{D60}
\end{eqnarray}
and $ F' = \partial F/ \partial z ,$ $ M_{V} = M_{V}^{(1)} + M_{V}^{(2)} .$
In the axisymmetric problem $ \partial {\bf B} / \partial \varphi = 0 .$
The thin-disk approximation implies that the spatial derivatives
of the mean magnetic field with respect to $ z $ are much larger than the
derivatives with respect to $ r .$
It is seen from Eqs. (\ref{D47})-(\ref{D49}) and (\ref{D58})-(\ref{D60})
that the contributions to the turbulent diffusion coefficients
$ \eta_{A}(B) $ and $ \eta_{B}(B) $ are from the tensor of
turbulent diffusion $ \eta_{ij}({\bf B}) ,$ the tensor $ \kappa_{ijk}({\bf B}) $
and the nonlinear velocity $ {\bf U}({\bf B}) + {\bf V}^{(N)}({\bf B}) .$ On the
other hand, contributions to the effective velocity $ V_{A}(B) $ are from
the tensor of turbulent diffusion $ \eta_{ij}({\bf B}) $
and the nonlinear velocity $ {\bf U}({\bf B}) + {\bf V}^{(N)}({\bf B}) .$
The functions $ \eta_{A}(B) ,$ $ \eta_{B}(B) $ and $ V_{A}(B) $ are given by
Eqs. (\ref{D61})-(\ref{D90}) in Appendix C.

The nonlinear dependencies: (A) of the turbulent magnetic
diffusion coefficients $ \eta_{A}(B) / \eta_{T}^{(v)} $ and $
\eta_{B}(B) / \eta_{T}^{(v)} ;$ (B) of the effective velocity $
V_{A}(B) / (B^{2})' ;$ and (C) of the nonlinear dynamo number $
D(B) / D_{\ast} $ are presented in Figures 1-3. Here $ D_{\ast} =
\alpha_{\ast} G h^{3} / \eta_{\ast}^{2} ,$ $ D(B) =
\alpha^{(v)}(B) G h^{3} / [\eta_{A}(B) \eta_{B}(B)] ,$ $
\eta_{\ast} = \eta_{T}^{(v)} + (2/3) \varepsilon_{\mu}^{(v)} ,$ $
\alpha_\ast $ is the maximum value of the hydrodynamic part of the
$ \alpha $ effect, $ h $ is the disc thickness and $
\alpha^{(v)}(B) = \alpha_{0}^{(v)} \Phi_{\alpha}(B) .$ For
simplicity we consider the case $ \epsilon = 0 .$

\begin{figure}
\centering
\includegraphics[width=8cm]{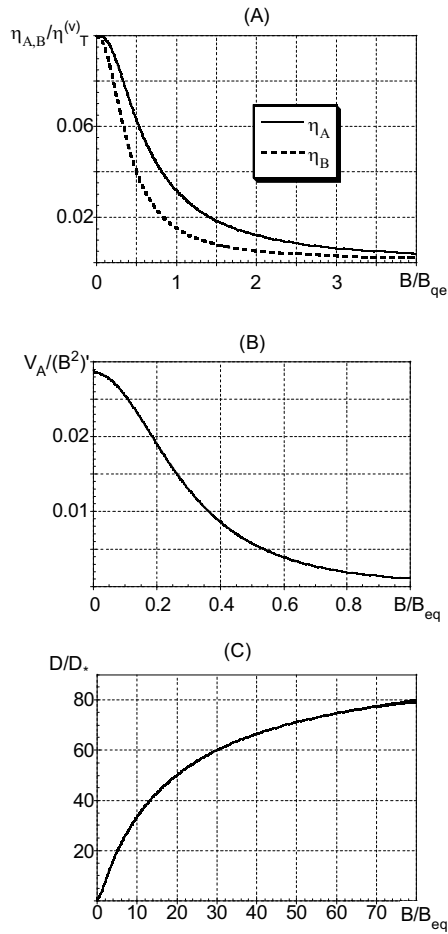}
\caption{\label{Fig1} (A). The nonlinear turbulent magnetic
diffusion coefficients; (B). The nonlinear effective velocity;
(C). The nonlinear dynamo number for $ \eta_{T}^{(h)} = 0 ;$ $
\varepsilon_{\mu}^{(v)} = - 1.35 \eta_{T}^{(v)}; $ $
\varepsilon_{\mu}^{(h)} = 0 .$}
\end{figure}

\begin{figure}
\centering
\includegraphics[width=8cm]{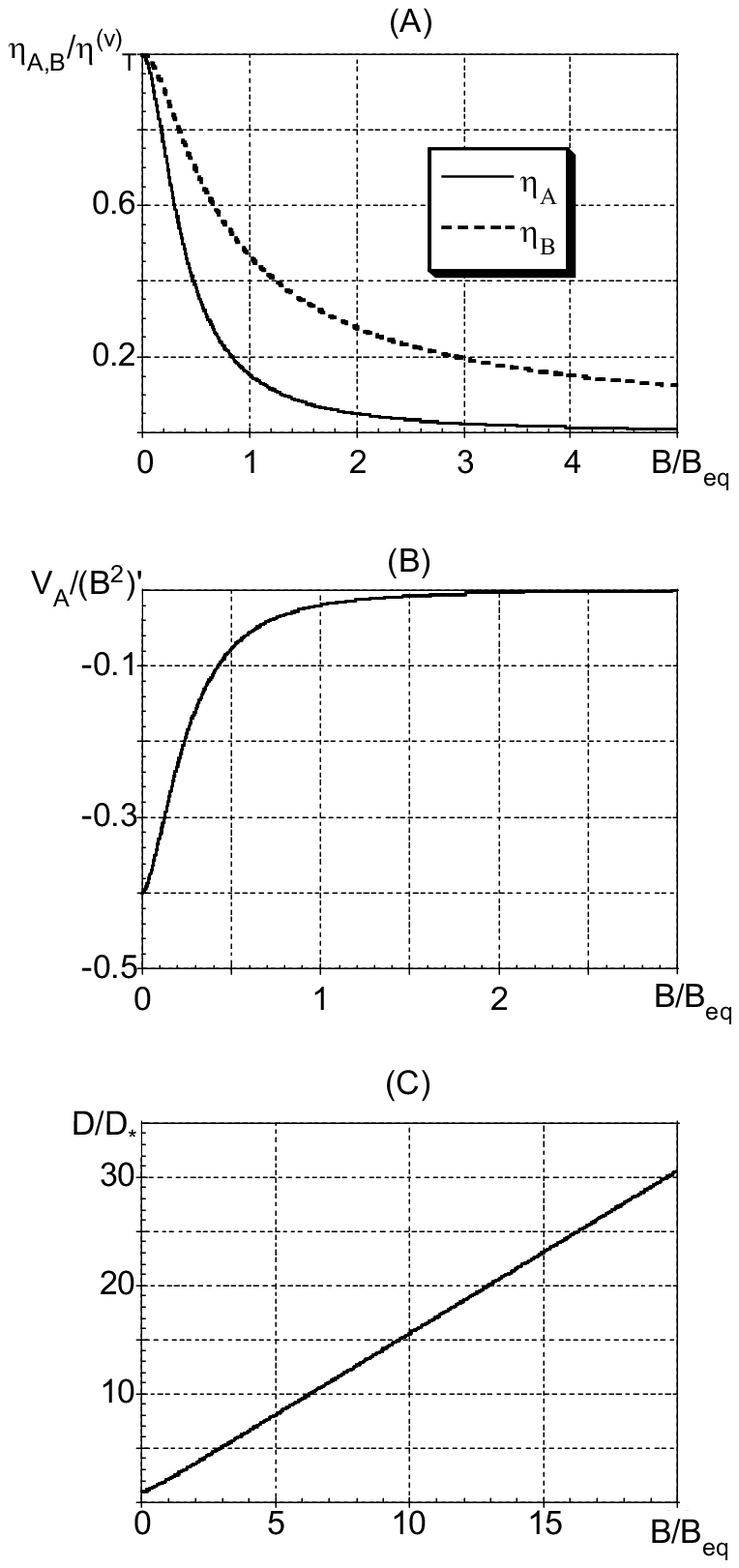}
\caption{\label{Fig2} (A). The nonlinear turbulent magnetic
diffusion coefficients; (B). The nonlinear effective velocity;
(C). The nonlinear dynamo number for $ \eta_{T}^{(v)} =
\eta_{T}^{(h)} ;$ $ \varepsilon_{\mu}^{(v)} =
\varepsilon_{\mu}^{(h)} = 0 .$}
\end{figure}

\begin{figure}
\centering
\includegraphics[width=8cm]{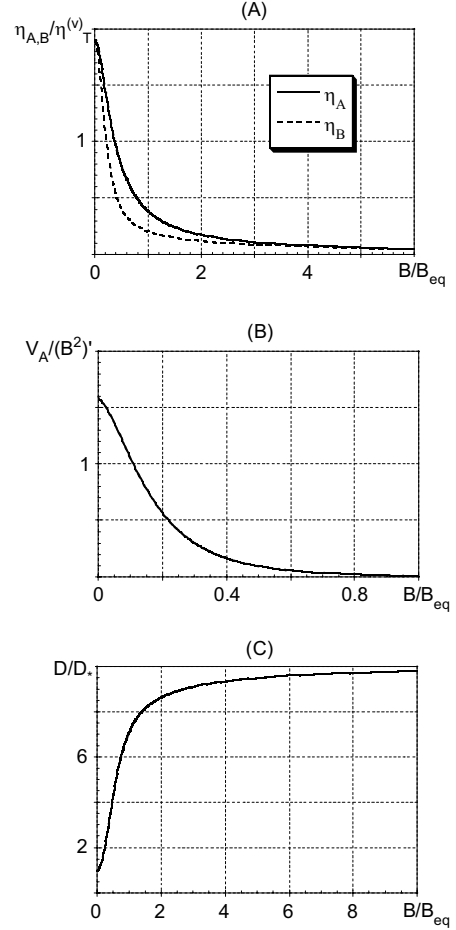}
\caption{\label{Fig3} (A). The nonlinear turbulent magnetic
diffusion coefficients; (B). The nonlinear effective velocity;
(C). The nonlinear dynamo number for $ \eta_{T}^{(h)} = 0 ;$ $
\varepsilon_{\mu}^{(v)} = 1.35 \eta_{T}^{(v)} ;$ $
\varepsilon_{\mu}^{(h)} = 0 .$}
\end{figure}

In order to separate the study of the algebraic and dynamic
nonlinearities we defined the nonlinear dynamo number $ D(B) $
using only the hydrodynamic part of the $ \alpha $ effect. We
considered three cases: two types of an anisotropic background
turbulence $ (\varepsilon_{\mu}^{(v)} = \pm 1.35 \eta_{T}^{(v)}; $
$ \varepsilon_{\mu}^{(h)} = 0) $ without magnetic fluctuations
(Fig. 1 and Fig. 3) and an isotropic $ (\varepsilon_{\mu}^{(v)} =
\varepsilon_{\mu}^{(h)} = 0) $ background turbulence with
equipartition of hydrodynamic and magnetic fluctuations (Fig. 2).
The negative degree of anisotropy $ \varepsilon_{\mu}^{(v)} $
implies that the vertical (along axis $ z )$ size of turbulent
elements is less than the horizontal size and positive $
\varepsilon_{\mu}^{(v)} $ means that the horizontal size is less
than the vertical size.

Figures 1-3 and the equations for $ \eta_{A}(B) $ and
$ \eta_{B}(B) $ show that the toroidal and poloidal magnetic
fields have different nonlinear turbulent magnetic diffusion coefficients.
In isotropic background turbulence (Fig. 2) the nonlinear effective velocity
$ V_{A}(B) $ is negative. The latter implies that it is diamagnetic velocity,
which results in  the field is pushed out from the regions with stronger mean
magnetic field. In the anisotropic background turbulence (Fig. 1 and Fig. 3)
the nonlinear effective velocity is positive, ({\em i.e.,} paramagnetic velocity
which causes the magnetic field tends to be concentrated in the regions with
stronger field). The sign of $ \varepsilon_{\mu}^{(v)} $ affects the
value of $ \eta_{A}(B) ,$ $ \eta_{B}(B) $ and $ V_{A}(B) ,$ {\em e.g.,}
for positive parameter of anisotropy the functions $ \eta_{A}(B) ,$
$ \eta_{B}(B) $ and $ V_{A}(B) $ are larger at least in one order of
magnitude then those for negative $ \varepsilon_{\mu}^{(v)} .$

The dependencies of the nonlinear dynamo number $ D(B) / D_{\ast} $
on the mean magnetic field $ B / B_{\rm eq} $ demonstrate that
the algebraic nonlinearity alone ({\em i.e.,} quenching of both, the nonlinear
$ \alpha $ effect and the nonlinear turbulent diffusion coefficients) cannot
saturate the growth of the mean magnetic field (where $ B_{\rm eq} =
\sqrt{\mu_{0} \rho} \, u_{0} ).$ Indeed, for anisotropic background
turbulence without magnetic fluctuations (Fig. 1 and Fig. 3) the nonlinear
dynamo number $ D(B) / D_{\ast} $ is a nonzero constant
for $ \beta \gg 1 ,$ {\em i.e.,} it is
independent on $ \beta .$ This is because for $ \beta \gg 1 $ the functions
$ \eta_{A} \propto 1 / \beta ,$ $ \eta_{B} \propto 1 / \beta $ and
$ \alpha \propto 1 / \beta^{2} $ [see Eqs. (\ref{R61})-(\ref{C10}) in
Appendix C]. In the case of isotropic background
turbulence with equipartition of hydrodynamic and magnetic fluctuations
(Fig. 2) the nonlinear dynamo number $ D(B) / D_{\ast} \propto \beta $ for
$ \beta \gg 1 $ because in this case the functions $ \eta_{A} \propto 1 / \beta^{2} ,$
$ \eta_{B} \propto 1 / \beta $ and $ \alpha \propto 1 / \beta^{2} $
[see Eqs. (\ref{R61})-(\ref{C10}) in Appendix C].
Note that the saturation of the growth of the mean magnetic field can be
achieved when the derivative of the nonlinear dynamo number
$ d D(B) / dB < 0 .$ Thus, the algebraic nonlinearity alone cannot
saturate the growth of the mean magnetic field.
We will show below that the combined effect of the algebraic and dynamic
nonlinearities can limit the growth of the mean magnetic field.

Equation (\ref{A4}) in nondimensional form is given by
\begin{eqnarray}
{\partial \alpha^{(h)}_{0} \over \partial t} + {\alpha^{(h)}_{0}
\over T} = 4 \biggl({h \over l_{0}} \biggr)^2 [ \eta_{B} B' A' -
(\eta_{A} A'' - V_{A} A'
\nonumber \\
+ \alpha B) B ] + [C |\alpha^{(v)}_{0}(z)| f_\eta(z)
\Phi_{\alpha}(B) \eta_{A}(B) B^2]' \;, \label{A5}
\end{eqnarray}
where $ C $ is a coefficient, $ f_\eta(z)$ describes the
inhomogeneity of the turbulent magnetic diffusion,
and we define $ f(z) = \alpha^{(v)}_{0}(z)
f_\eta(z) .$ We use here the standard dimensionless form of the galactic
dynamo equation (see, {\em e.g.,} \cite{RSS88}), in
particular, the length is measured in units of the disc thickness $h$,
the time is measured in units of $ h^{2} / \eta_{T}^{(v)} $ and $B$ is
measured in units of the equipartition energy $B_{\rm eq} =
\sqrt{\mu_{0} \rho} \, u_{0} $.  Here $u_{0}$ is the characteristic turbulent
velocity in the maximum scale $l_{0}$ of turbulent motions,
$ \eta_{T}^{(v)} = l _{0} u _{0} / 3 $ and $ \alpha^{(v)}_{0} ,$ $ \alpha^{(h)}_{0} $
and $ \alpha $ are measured
in units of $ \alpha_\ast $ (the maximum value of the hydrodynamic
part of the $ \alpha $ effect). For galaxies $ h / l_{0} \sim 5 $ and
$ C \sim 0.05 - 0.1 .$ Nondimensional equations for $ A $ and $
B $ are given by
\begin{eqnarray}
\partial B / \partial t &=& (\eta_{B} B')' - D_{0} A' \;,
\label{A6} \\
\partial A / \partial t &=& \eta_{A} A'' - V_{A} A' + \alpha B
\;,
\label{A7}
\end{eqnarray}
where $ D_{0} = \alpha_{\ast} G h^{3} / \eta_{T}^{2} $ and $ B_{r} = - A'(z) .$
In a steady state Eqs. (\ref{A5})-(\ref{A7}) yield
\begin{eqnarray}
[\eta_{B}(B) B']^{2} + 2 C D_{0} \Phi_{\alpha}(B) \eta_{A}(B) B^2 |f(z)|
= 0 \;,
\label{AA8}
\end{eqnarray}
where we used the following boundary conditions
$ B(z=\pm 1) = 0 ,$ $ B'(z=0) = 0 $ and $ f(z=0) = 0 .$
The solution of Eq. (\ref{AA8}) for negative $ D_{0} $
is given by
\begin{eqnarray}
\int_{0}^{B} \chi(\tilde B) \,d \tilde B = \sqrt{2 C |D_{0}|}
\int_{|z|}^{1} \sqrt{f(\tilde z)} \,d \tilde z \;,
\label{A8}
\end{eqnarray}
where $ \chi(B) = \eta_{B}(B) / [\Phi_{\alpha}(B) \eta_{A}(B) B^2]^{1/2} .$
Consider the case $ \epsilon = 0 .$
For $ \beta \gg 1 $ ({\em i.e.,} for  $ B \gg 1/ \sqrt{8}) $ the equilibrium
mean toroidal magnetic field $ B(z) $ is given by
\begin{eqnarray}
B(z) \approx \sqrt{2} C |D_{0}| \biggl( \int_{|z|}^{1} \sqrt{f(\tilde z)} \,d
\tilde z \biggr)^{2} \;,
\label{A9}
\end{eqnarray}
where we used that for $ \beta \gg 1 $ the functions $ \eta_{A}(B) \sim 3 / 5 \beta ,$
$ \eta_{B}(B) \sim 2 / 5 \beta ,$ $ \Phi_{\alpha}(B) \sim 2 / \beta^{2} ,$
and $ \chi(B) \sim 2 / \sqrt{\beta} .$ Here for simplicity we
considered the case $ \varepsilon_{\mu}^{(v)} = 0 .$
In a steady state $ A(z) = - \eta_{B}(B) B' / |D_{0}| .$
Now we specify the profile of the function $ f(z) ,$ {\em e.g.,}
$ f(z) = f_{\ast}[\sin(\pi z / 2)]^{2k+1} [\cos(\pi z / 2)]^{2} ,$
where $ k = 1; 2; 3; \ldots $ and
\begin{eqnarray*}
f_{\ast} = \biggl({2 k + 3 \over 2} \biggr) \biggl({2 k + 3 \over
2 k + 1} \biggr)^{(2k+1)/2} \; .
\end{eqnarray*}
The function $ f(z) $ changes in the
interval $ 0 \leq f(z) \leq 1 $ and it has a maximum
$ f(z=z_{m}) = 1 $ at $ z_{m} = (2/\pi) \arctan[\sqrt{(2k+1)/2}] .$ Equation
(\ref{A9}) for this profile $ f(z) $ with $ k = 2 $ yields
\begin{eqnarray}
B(z) \approx 0.4 C |D_{0}| \{1 - [\sin(\pi z / 2)]^{7/2} \}^{2} \; .
\label{A11}
\end{eqnarray}
Equation (\ref{A11}) describes the equilibrium configuration of the
mean toroidal magnetic field. Thus, the saturation of the growth of the
mean magnetic field is caused by both, the algebraic and dynamic
nonlinearities. The dynamic nonlinearity is determined by the dynamic
equation (\ref{A5}) whereas the algebraic nonlinearity implies
the nonlinear dependencies of the turbulent magnetic diffusion coefficients
$ \eta_{A}(B) $ and $ \eta_{B}(B) $ and of the effective velocity
$ V_{A}(B) $ on the mean magnetic field [see Eqs. (\ref{D61})-(\ref{D90})].

\section{Discussion}

In this study we calculated the nonlinear tensor of
turbulent magnetic diffusion, the nonlinear $ \kappa $-tensor,
the nonlinear effective velocity, and other coefficients defining the
mean electromotive force for an anisotropic turbulence.
The obtained results were specified for an anisotropic background turbulence
with one preferential direction. We found that the turbulent magnetic
diffusion coefficients for the toroidal and poloidal magnetic fields
are different. We demonstrated that even for a homogeneous turbulence there is
the nonlinear effective velocity which can be a diamagnetic or paramagnetic
velocity depending on anisotropy of turbulence and level of magnetic
fluctuations in the background turbulence. The diamagnetic velocity implies
that the field is pushed out from the regions with stronger mean magnetic
field, while the paramagnetic velocity causes the magnetic field tends
to be concentrated in the regions with stronger field.

Note that dependencies of the $ \alpha $-effect, the turbulent
magnetic diffusion coefficient and the effective drift velocity on
the mean magnetic field for an isotropic turbulence have been found in
\cite{K91,RK93,KPR94} using a modified second order correlation
approximation. Our results are different from that obtained in
\cite{K91,RK93,KPR94}. The reason is that in \cite{K91,RK93,KPR94}
a phenomenological procedure was used. In particular, in the first
step of the calculations the nonlinear terms in the magnetohydrodynamic
equations were dropped (which is valid for small hydrodynamic and
magnetic Reynolds numbers or in a highly conductivity limit and
small Strouhle numbers). In the next step of the calculations
in \cite{K91,RK93,KPR94} it was assumed that $ \nu = \eta =
l_{c}^{2} / \tau_{c} ,$ where $ l_{c} $ and $ \tau_{c} $ are the
correlation length and time of turbulent velocity field.
The latter is valid when the hydrodynamic and magnetic Reynolds numbers
are of the order of unit.
In the present paper we use a different procedure (the $ \tau
$-approximation) for large hydrodynamic and magnetic Reynolds numbers.

In this study we also demonstrated an important role of two types of
nonlinearities (algebraic and dynamic) in the mean-field dynamo.
The algebraic nonlinearity is determined by a nonlinear
dependence of the mean electromotive force on the mean magnetic field.
The dynamic nonlinearity is determined by a differential equation for the
magnetic part of the $ \alpha $-effect. This equation is a consequence
of the conservation of the total magnetic helicity (which includes
both, the magnetic helicity of the mean magnetic field and the
magnetic helicity of small-scale magnetic fluctuations). We found that
at least for the $ \alpha \Omega $ axisymmetric dynamo the algebraic nonlinearity
alone [{\em i.e.,} the nonlinear functions $ \alpha(B) ,$ $ \eta_{A}(B) ,$
$ \eta_{B}(B) $ and $ V_{A}(B) ]$
cannot saturate the dynamo generated mean magnetic field.
The important parameter which characterizes the algebraic nonlinearity
is the nonlinear dynamo number $ D(B) .$ The saturation of the growth of
the dynamo generated mean magnetic field by the algebraic nonlinearity
alone is possible when the derivative $ d D(B) / dB < 0 .$
We found that for the $ \alpha \Omega $ axisymmetric dynamo the nonlinear dynamo
number $ D(B) $ is either a constant or $ D(B) \propto B $ for $ B >
B_{\rm eq} / 3 $ depending on the model of the background turbulence.
Therefore, in this case the algebraic nonlinearity
alone cannot saturate the dynamo generated mean magnetic field.

The situation is changed when the dynamic nonlinearity is taken into
account. The crucial point is that the dynamic equation for the
magnetic part of the $ \alpha $-effect ({\em i.e.,}
the dynamic nonlinearity) includes the flux of the
magnetic helicity. Without the flux, the total magnetic helicity is
conserved locally and the level of the saturated mean magnetic field
is very low \cite{KMRS2000}. The flux of the magnetic helicity results in that
the total magnetic helicity is not conserved locally because the
magnetic helicity of small-scale magnetic fluctuations is
redistributed by a helicity flux. In this case
an integral of the total magnetic helicity over the disc is conserved.
The equilibrium state is given by a balance between magnetic helicity
production and magnetic helicity transport \cite{KMRS2000}. These two types of the
nonlinearities (algebraic and dynamic) results in
the equilibrium strength of the mean magnetic field
is of order that of the equipartition field $ B_{\rm eq} $
(see Section V) in agreement with observations of the galactic magnetic fields
(see, {\em e.g.,} \cite{RSS88}).

\begin{acknowledgments}
We have benefited from stimulating discussions on nonlinear dynamo
with A. Brandenburg, D. Moss, K.-H. R\"{a}dler, P. H. Roberts, A.
Ruzmaikin, D. Sokoloff and E. T. Vishniac. We are also grateful to
the anonymous referee for very useful and important comments which
strongly improved our paper. This work was partially supported by
INTAS Program Foundation (Grant No. 99-348).
\end{acknowledgments}

\appendix

\section{Calculation of the mean electromotive force}

Let us derive equations for the second moments. For this purpose we rewrite
Eqs. (\ref{Q4}) and (\ref{Q5}) in a Fourier space
and repeat twice the vector multiplication of Eq. (\ref{Q4}) by
the wave vector $ {\bf k} $ . The result is given by
\begin{eqnarray}
{du_i({\bf k},t) \over dt} & = & {1 \over \mu_{0} \rho} [(2
P_{ip}(k) - \delta_{ip}) \hat S_{p}^{(c)}(b;B) + \hat
S_{i}^{(b)}(b;B)]
\nonumber \\
& & - {\tilde T}_i - \nu k^2 u_i - \tilde F_{i} \;,
\label{Q1}\\
{db_i({\bf k},t) \over dt} & = & \hat S_{i}^{(b)}(u;B) - \hat
S_{i}^{(c)}(u;B) + G_i - \eta k^2 b_i \;, \label{Q2}
\end{eqnarray}
where $ \hat S_{i}^{(c)}(a;A) = i \int a_p ({\bf k}-{\bf Q}) Q_{p}
A_i ({\bf Q}) \,d {\bf Q} \;, $  $ \hat S_{i}^{(b)}(a;A) = i k_{p}
\int a_i ({\bf k}-{\bf Q}) A_p ({\bf Q}) \,d {\bf Q} \;, $
$ {\bf \tilde T} ={\bf k} {\bf \times} ({\bf k}{\bf \times}{\bf T}) /k^2 \;,$
$ {\bf \tilde F} ({\bf k},{\bf R},t) = {\bf k} {\bf \times} ({\bf k}
{\bf \times} {\bf F} ({\bf k},{\bf R})) / k^2 \rho \;,$
$ P_{ij}(k) =  \delta _{ij} - k_{ij} ,$
$ \delta_{ij} $ is the Kronecker tensor and $ k_{ij} = k_i  k_j / k^2 .$
We use the two-scale approach, {\em i.e.,} a correlation function
\begin{eqnarray*}
\langle u_i ({\bf x} ) u_j ({\bf  y}) \rangle = \int \langle u_i
({\bf  k}_1) u_j ({\bf k}_2) \rangle  \exp \{i( {\bf  k}_1 {\bf
\cdot} {\bf x}
\\
+ {\bf  k}_2 {\bf \cdot} {\bf y}) \} \,d{\bf k}_1 \, d{\bf  k}_2
\\
= \int f_{ij}( {\bf k, K} ) \exp{(i {\bf k} {\bf \cdot} {\bf r} +
i {\bf K} {\bf \cdot} {\bf R}) } \,d {\bf  k} \,d {\bf  K} \;,
\\
= \int f_{ij}( {\bf k, R} ) \exp{(i {\bf k} {\bf \cdot} {\bf r}) }
\,d {\bf  k} \;,
\\
f_{ij}({\bf k, R} ) = \int \langle u_i ({\bf k} + {\bf  K} / 2 )
u_j( -{\bf k} + {\bf  K}  / 2 ) \rangle
\\
= \times \exp{(i {\bf K} {\bf \cdot} {\bf R}) } \,d {\bf  K} \;,
\end{eqnarray*}
where $ {\bf R} = ( {\bf x} +  {\bf y}) / 2  , \quad {\bf r} =
{\bf x} - {\bf y}, \quad {\bf K} = {\bf k}_1 + {\bf k}_2, \quad
{\bf k} = ({\bf k}_1 - {\bf k}_2) / 2 ,$ $ {\bf R} $ and $ {\bf K}
$  correspond to the large scales, and $ {\bf r} $ and $ {\bf k} $
to the  small ones (see, {\em e.g.,}  \cite{RS75,KR94}). The
others second moments have the same form, {\em e.g.,}
\begin{eqnarray*}
h_{ij}({\bf k, R}) &=& \int \langle b_i ({\bf k} + {\bf  K} / 2)
b_j( -{\bf k} + {\bf  K}  / 2 ) \rangle
\nonumber\\
& & \times \exp{(i {\bf K} {\bf \cdot} {\bf R}) } \,d {\bf  K} /
\mu_{0} \rho \;,
\\
g_{ij}({\bf k, R}) &=& \int \langle b_i ({\bf k} + {\bf  K} / 2)
u_j( -{\bf k} + {\bf  K}  / 2 ) \rangle
\nonumber\\
& & \times \exp{(i {\bf K} {\bf \cdot} {\bf R}) } \,d {\bf  K}
\; .
\end{eqnarray*}
The two-scale approach is valid when $ (1 / B) (d B / dR)
\ll l_{0}^{-1} ,$ where $ B = \vert {\bf B} \vert .$
Now we derive the equations for the correlation functions
$ f_{ij}({\bf k , R}) ,$ and $ h_{ij}({\bf k , R}) ,$
and $ g_{ij}({\bf k , R}) $
\begin{eqnarray}
\partial f_{ij} / \partial t & = & i({\bf k } {\bf \cdot} {\bf B}) \Phi_{ij}
+ M_{ij} + F_{ij} - 2 \nu k^2 f_{ij} \;,
\label{R3} \\
\partial h_{ij} / \partial t & = & - i({\bf k }{\bf \cdot} {\bf B}) \Phi_{ij} +
R_{ij} - 2 \eta k^2 h_{ij} \;,
\label{R4} \\
\partial g_{ij} / \partial t & = & I_{ij} + C_{ij}
- (\nu + \eta) k^2 g_{ij} \;,
\label{Q3} \\
I_{ij} &=& i({\bf k} {\bf \cdot} {\bf B}) (f_{ij} -  h_{ij}) +
(1/2) ({\bf B} {\bf \cdot} \bec{\nabla}) (f_{ij} + h_{ij})
\nonumber \\
& & - f_{pj} B_{i,p} + h_{ip} (2 P_{jl}(k)
- \delta_{jl}) B_{l,p}
\nonumber\\
& & - B_{p,q} k_{p}(f_{ijq} + h_{ijq}) \;, \label{T23}
\end{eqnarray}
where $ \bec{\nabla} = \partial / \partial {\bf R} ,$ $ f_{ijq} = (1/2) \partial
f_{ij} / \partial k_{q} ,$ $ h_{ijq} = (1/2) \partial h_{ij} / \partial k_{q} ,$
and $ F_{ij}({\bf k},{\bf R}) = \langle \tilde F_i ({\bf k},{\bf R})
u_j(-{\bf k},{\bf R}) \rangle + \langle u_i({\bf k},{\bf R})
\tilde F_j(-{\bf k },{\bf R})\rangle \;,$ $ \, B_{i,j} = \partial B_{i}
/ \partial R_{j} ,$ and
\begin{eqnarray}
\Phi_{ij}({\bf k },{\bf R}) = [g_{ij}({\bf k},{\bf R})
- g_{ji}(-{\bf k},{\bf R})] / \mu_{0} \rho \; .
\label{T40}
\end{eqnarray}
The third moments are given by $ M_{ij}({\bf k},{\bf R}) = - \langle \tilde
T_i ({\bf k }) u_j(-{\bf k}) \rangle - \langle u_i({\bf k}) \tilde
T_j(-{\bf k}) \rangle ,$ $ R_{ij}({\bf k},{\bf R}) = \langle \tilde
G_i ({\bf k }) b_j(-{\bf k}) \rangle + \langle b_i({\bf k}) \tilde
G_j(-{\bf k}) \rangle $ and $ C_{ij}({\bf k},{\bf R}) = \langle \tilde
G_i ({\bf k }) u_j(-{\bf k}) \rangle - \langle b_i({\bf k}) \tilde
T_j(-{\bf k}) \rangle .$

For the derivation of Eqs. (\ref{R3})-(\ref{T23}) we performed several
calculations that are similar to the following, which arose in
computing $ \partial g_{ij} / \partial t .$ The other calculations
follow similar lines and are not given here. Let us define
$ Y_{ij}({\bf k, R}) $ by
\begin{eqnarray*}
&& Y_{ij}({\bf k, R}) = \int \langle \hat S^{(b)}_i (u;B; {\bf k}
+ {\bf K} / 2) u_j(-{\bf k} + {\bf K} / 2) \rangle
\\
&& \times \exp{(i {\bf K} {\bf \cdot} {\bf R}) } \,d {\bf  K} = i
\int (k_{p} + K_{p}/2) B_{p}({\bf  Q}) \exp(i {\bf K} {\bf \cdot}
{\bf R})
\\
&& \times \langle u_i ({\bf k} + {\bf  K} / 2 - {\bf  Q})
u_j(-{\bf k} + {\bf  K}  / 2) \rangle \,d {\bf  K} \,d {\bf  Q}
\; .
\end{eqnarray*}
Next, we introduce new variables: $ \tilde {\bf k}_{1} =
{\bf k} + {\bf  K} / 2 - {\bf  Q} ,$ $ \tilde {\bf k}_{2} =
- {\bf k} + {\bf  K} / 2 $ and $ \tilde {\bf k} = (\tilde {\bf k}_{1} -
\tilde {\bf k}_{2}) / 2 = {\bf k} - {\bf  Q} / 2,$ $ \tilde {\bf K} =
\tilde {\bf k}_{1} + \tilde {\bf k}_{2} = {\bf  K} - {\bf  Q} .$
Therefore,
\begin{eqnarray}
Y_{ij}({\bf k, R} ) &=& i \int f_{ij}({\bf k} - {\bf  Q} / 2, {\bf
K} - {\bf  Q}) (k_{p} + K_{p}/2) B_{p}({\bf  Q})
\nonumber\\
& & \times \exp{(i {\bf K} {\bf \cdot} {\bf R})} \,d {\bf  K} \,d
{\bf  Q} \; . \label{NNN3}
\end{eqnarray}
Since $ |{\bf Q}| \ll |{\bf k}| $ we use the Taylor expansion
\begin{eqnarray}
f_{ij}({\bf k} - {\bf Q}/2, {\bf  K} - {\bf  Q}) \simeq
f_{ij}({\bf k},{\bf  K} - {\bf  Q})
\nonumber\\
- \frac{1}{2} {\partial f_{ij}({\bf k},{\bf  K} - {\bf Q}) \over
\partial k_s} Q_s  + O({\bf Q}^2) \;, \label{NNN1}
\end{eqnarray}
and the following identities:
\begin{eqnarray}
&& [f_{ij}({\bf k},{\bf R}) B_{p}({\bf R})]_{\bf  K} = \int
f_{ij}({\bf k},{\bf  K} - {\bf  Q}) B_{p}({\bf Q}) \,d {\bf  Q}
\;,
\nonumber \\
&& \nabla_{p} [f_{ij}({\bf k},{\bf R}) B_{p}({\bf R})] = \int i
K_{p} [f_{ij}({\bf k},{\bf R}) B_{p}({\bf R})]_{\bf  K}
\nonumber\\
&& \times \exp{(i {\bf K} {\bf \cdot} {\bf R})} \,d {\bf  K} \; .
\label{NNN2}
\end{eqnarray}
Therefore, Eqs. (\ref{NNN3})-(\ref{NNN2}) yield
\begin{eqnarray}
Y_{ij}({\bf k},{\bf R}) &\simeq& [i({\bf k } \cdot {\bf B}) +
(1/2) ({\bf B} \cdot \bec{\nabla})] f_{ij}({\bf k},{\bf R})
\nonumber\\
& & - k_{p} f_{ijs}({\bf k},{\bf R}) B_{p,s} \; . \label{NNN4}
\end{eqnarray}

In Eqs. (\ref{R3}) and (\ref{R4}) we neglected the
terms $ \propto ({\bf B} {\bf \cdot} \bec{\nabla}) g_{ij} $ and
$ \propto B_{i,p} g_{pj} $ because they contribute to the modification of
the mean Lorentz force by the turbulence effect (see, {\em e.g.,}
\cite{KRR90,KMR96}). In Eq. (\ref{Q3}) we neglected the second and higher
derivatives over $ {\bf R} .$ We also neglected in Eq. (\ref{Q3}) the terms
which are of the order of
$ {\rm Rm}^{-1} \bec{\nabla} (B_{i},f_{ij},h_{ij}) $
and $ {\rm Re}^{-1} \bec{\nabla} (B_{i};f_{ij};h_{ij}) .$
When the mean magnetic field is zero Eq. (\ref{R3}) reads
\begin{eqnarray}
\partial f_{ij}^{(0)} / \partial t = M_{ij}^{(0)} + F_{ij}^{(0)} -
2 \nu k^2 f_{ij}^{(0)} \;,
\label{PPR3}
\end{eqnarray}
We assume that $ F_{ij} $ is not changed during the generation
of the mean magnetic field, {\em i.e.,} $ F_{ij} = F_{ij}^{(0)} .$
This implies an assumption of a constant power
of the source of turbulence.

Now we split all correlation functions ({\em i.e.,} $ f_{ij}, h_{ij},
g_{ij}, \Phi_{ij}) $ into two parts, {\em e.g.,} $ f_{ij} = f_{ij}^{(N)} +
f_{ij}^{(S)} ,$ where $ f_{ij}^{(N)} =
[f_{ij}({\bf k},{\bf R}) + f_{ij}(-{\bf k},{\bf R})] / 2 $ and
$ f_{ij}^{(S)} = [f_{ij}({\bf k},{\bf R}) - f_{ij}(-{\bf k},{\bf R})] /
2 .$ Next, we use $ \tau $ approximation which is determined by
Eqs. (\ref{A1})-(\ref{A3}). We assume that $ \eta k^2 \ll \tau ^{-1} $ and
$ \nu k^2 \ll \tau ^{-1} $ for the inertial range of turbulent fluid flow.
We also  assume that the characteristic time of variation of the mean magnetic
field $ {\bf B} $ is substantially longer than the correlation time
$ \tau(k) $ for all turbulence scales. Thus, Eqs. (\ref{R3})-(\ref{Q3})
yield
\begin{eqnarray}
f_{ij}^{(N)} &\approx& f_{ij}^{(0N)} + i \tau ({\bf k} {\bf \cdot} {\bf B})
\Phi_{ij}^{(S)} \;,
\label{Q39}\\
h_{ij}^{(N)} &\approx& h_{ij}^{(0N)} - i \tau ({\bf k} {\bf \cdot} {\bf B})
\Phi_{ij}^{(S)} \;,
\label{Q8}\\
f_{ij}^{(S)} &\approx& f_{ij}^{(0S)} + i \tau ({\bf k} {\bf \cdot} {\bf B})
\Phi_{ij}^{(N)} \;,
\label{Q39Q}\\
g_{ij} &\approx& \tau I_{ij} \;,
\label{T41}
\end{eqnarray}
where $ \psi = 2 ({\bf k} {\bf \cdot} {\bf B} \tau)^2 / \mu_{0} \rho ,$
$ k_{ij} = k_{i} k_{j} / k^{2} ,$ $ f_{ij}^{(0N)} $ and $
f_{ij}^{(0S)} $ describe the nonhelical and helical tensors of the background
turbulence. The tensor $ h_{ij}^{(S)} $ is
determined by an evolutionary equation
(see, {\rm e.g.,} \cite{ZRS83,KR82,VK83,KRR94,GD94,KR99,KMRS2000} and
Section III-C). Now we calculate $ \Phi_{ij}^{(N)} $ and $ \Phi_{ij}^{(S)} .$
The definition of $ \Phi_{ij} ,$ given by Eq. (\ref{T40}), and Eq.
(\ref{T41}) yield
\begin{eqnarray}
\Phi_{ij}({\bf k},{\bf R}) \approx \tau (\mu_{0} \rho)^{-1}
[I_{ij}({\bf k},{\bf R}) - I_{ji}(-{\bf k},{\bf R})]  \; .
\label{T42}
\end{eqnarray}
Substituting Eq. (\ref{T23}) into Eq. (\ref{T42}) and using Eqs.
(\ref{T10}) and (\ref{T11}) we obtain
\begin{eqnarray}
\Phi_{ij} \approx {\tau \over \mu_{0} \rho} \{2 i({\bf k} {\bf
\cdot} {\bf B}) (f_{ij} - h_{ij}) - B_{i,p} (f_{pj} + h_{pj})
\nonumber \\
+ B_{j,p} (f_{ip} + h_{ip}) + 2 B_{l,p} (h_{pj} k_{li} - h_{ip}
k_{lj}) \} \; . \label{T43}
\end{eqnarray}
Now using Eqs. (\ref{Q39})-(\ref{Q39Q}) and (\ref{T43}) we get
\begin{eqnarray}
\Phi_{ij}^{(N)} &\approx& {\tau \over (1 + \psi) \mu_{0} \rho} \{2
i({\bf k} {\bf \cdot} {\bf B}) (f_{ij}^{(0S)} - h_{ij}^{(S)})
\nonumber \\
& & + B_{j,p} (f_{pi}^{(0N)} + h_{pi}^{(0N)}) - B_{i,p}
(f_{pj}^{(0N)} + h_{pj}^{(0N)})
\nonumber \\
& & + 2 B_{p,l} (h_{jl}^{(N)} k_{pi} - h_{il}^{(N)} k_{pj}) \} \;,
\label{Q9} \\
\Phi_{ij}^{(S)} &\approx& {2 i \tau ({\bf k} {\bf \cdot} {\bf B})
\over (1 + 2 \psi) \mu_{0} \rho}  (f_{ij}^{(0N)} - h_{ij}^{(0N)})
\nonumber \\
& & + O(B_{i,j}) \; . \label{Q70}
\end{eqnarray}
To calculate the electromotive force we do not take into account the second and
higher orders spatial derivatives of the  mean magnetic field.
This implies that in the tensor $ b_{ijk} $ (and, therefore, in
the tensor $ \Phi_{ij}^{(S)}) $ we neglect the first and higher orders
spatial derivatives of the mean magnetic field [see Eqs. (\ref{P20}),
(\ref{Q70}) and Eq. (\ref{T45}) below].

Note that Eqs. (\ref{Q39}) and (\ref{Q8}) yield
\begin{eqnarray}
f_{ij}^{(N)} + h_{ij}^{(N)} \approx f_{ij}^{(0N)} + h_{ij}^{(0N)} \; .
\label{T44}
\end{eqnarray}
This is in agreement with that a uniform mean magnetic
field performs no work on the turbulence. It can only redistribute
the energy between hydrodynamic fluctuations and magnetic fluctuation.
Analysis in \cite{KMR96} showed that a change of the total
energy (kinetic and magnetic) of fluctuations caused by a nonuniform
mean magnetic field is of the order $ \sim \tau \eta^{(v)}_{T} \Delta
B^{2} .$ In the case of a very strong mean magnetic field Eq. (\ref{T44})
can be violated.

Using Eqs. (\ref{Q39})-(\ref{Q39Q}) and (\ref{Q9})-(\ref{Q70}) we calculate
the electromotive force
$ {\cal E}_{i}({\bf r}=0) = \int {\cal E}_{i}({\bf k}) \,d {\bf k} ,$
where the Fourier component $ {\cal E}_{i}({\bf k}) = (\mu_{0} \rho / 2)
\varepsilon_{imn} \Phi_{nm}^{(N)}({\bf k}) ,$ and $ \varepsilon_{ijk} $ is
the Levi-Civita tensor. The electromotive force is given by
Eqs. (\ref{P20})-(\ref{P22}).
Substituting Eq. (\ref{Q8}) into Eq. (\ref{P22}) we get
\begin{eqnarray}
b_{ijk} = \int \tau (1 + \psi)^{-1} \{ \varepsilon_{ijn}
(f_{kn}^{(0N)} + h_{kn}^{(0N)})
\nonumber \\
- 2 \varepsilon_{imn} k_{mj} [h_{nk}^{(0N)} - i \tau ({\bf k} {\bf
\cdot} {\bf B}) \Phi_{nk}^{(S)}] \} \,d {\bf k} \; . \label{T45}
\end{eqnarray}
The integration in $ {\bf k} $-space in Eq. (\ref{T45}) yields
\begin{eqnarray}
b_{ijk} = \varepsilon_{ijn} \lambda^{(P)}_{nk}(\beta) + 2 \varepsilon_{inm}
\zeta^{(C)}_{nkjm}(\beta) \; .
\label{C20}
\end{eqnarray}
Hereafter we use the following definitions:
\begin{eqnarray}
X^{(C)}_{ijk\ldots}(\beta) &=& X^{(v)}_{ijk\ldots}(\beta) -
X^{(v)}_{ijk\ldots}(\sqrt{2}\beta)
\nonumber \\
& & + X^{(h)}_{ijk\ldots}(\sqrt{2}\beta) \;,
\label{DD1} \\
X^{(M)}_{ijk\ldots}(\beta) &=& X^{(v)}_{ijk\ldots}(\beta) -
X^{(h)}_{ijk\ldots}(\beta)  \;,
\label{DD2} \\
X^{(P)}_{ijk\ldots}(\beta) &=& X^{(v)}_{ijk\ldots}(\beta) +
X^{(h)}_{ijk\ldots}(\beta)  \;,
\label{DD3}
\end{eqnarray}
and
\begin{eqnarray}
\lambda^{(a)}_{ij}(\beta) &=& \int {c_{ij}({\bf k}) \tau(k) \over
1 + \psi(\beta,{\bf k})} \,d{\bf k} \;,
\label{C23} \\
\zeta^{(a)}_{ijmn}(\beta) &=& \int {c_{ij}({\bf k}) \tau(k) \over
1 + \psi(\beta,{\bf k})} k_{mn} \,d{\bf k} \;,
\label{C21}
\end{eqnarray}
and $ \beta_{i} = 4 B_{i} / (u_{0} \sqrt{2 \mu_{0} \rho}) ,$
$ \quad \psi(\beta,{\bf k}) = [(\bec{\beta} {\bf \cdot} {\bf k}) u_{0} \tau / 2]^{2}
,$ $ \quad c_{ij} = f_{ij}^{(0N)} $ for $ \lambda^{(v)}_{ij} $ and
$ c_{ij} = h_{ij}^{(0N)} $ for $ \lambda^{(h)}_{ij} .$

For the calculation of the tensor
$ b_{ijk} $ we specified a model of the background turbulence
({\em i.e.,} turbulence with zero mean magnetic field). The turbulent velocity
and magnetic fields of the background turbulence are determined
by Eq. (\ref{Q19}). To integrate over the angles in $ {\bf k} $--space in
Eqs. (\ref{C23}) and (\ref{C21}) we use the following identities:
\begin{eqnarray}
&&\int {k_{ij} \sin \theta \over 1 + a \cos^{2} \theta} \,d\theta
\,d\varphi = \bar A_{1} \delta_{ij} + \bar A_{2} \beta_{ij} \;,
\label{C22} \\
&& \int {k_{ijmn} \sin \theta \over 1 + a \cos^{2} \theta}
\,d\theta \,d\varphi = \bar C_{1} (\delta_{ij} \delta_{mn} +
\delta_{im} \delta_{jn} + \delta_{in} \delta_{jm})
\nonumber \\
&& + \bar C_{2} \beta_{ijmn} + \bar C_{3} (\delta_{ij} \beta_{mn}
+ \delta_{im} \beta_{jn} + \delta_{in} \beta_{jm} + \delta_{jm}
\beta_{in}
\nonumber \\
&&+ \delta_{jn} \beta_{im} + \delta_{mn} \beta_{ij}) \;,
\label{C24}
\end{eqnarray}
where $ a = [\beta u_{0} k \tau(k) / 2]^{2} ,$
$ \hat \beta_{i} =  \beta_{i} / \beta ,$
$ \beta_{ij} = \hat \beta_{i} \hat \beta_{j} ,$
$ \beta_{ijk..} = \hat \beta_{i} \hat \beta_{j} \hat \beta_{k} \ldots ,$
and
\begin{eqnarray*}
\bar A_{1} &=& {2 \pi \over a} \biggl[(a + 1) {\arctan (\sqrt{a}) \over
\sqrt{a}} - 1 \biggr] \;,
\\
\bar A_{2} &=& - {2 \pi \over a} \biggl[(a + 3) {\arctan (\sqrt{a}) \over
\sqrt{a}} - 3 \biggr] \;,
\\
\bar C_{1} &=& {\pi \over 2a^{2}} \biggl[(a + 1)^{2} {\arctan (\sqrt{a})
\over \sqrt{a}} - {5 a \over 3} - 1 \biggr]  \;,
\\
\bar C_{2} &=& {\pi \over 2a^{2}} \biggl[(3 a^{2} + 30 a + 35)
{\arctan (\sqrt{a}) \over \sqrt{a}} - {55 a \over 3} - 35 \biggr]  \;,
\\
\bar C_{3} &=& - {\pi \over 2a^{2}} \biggl[(a^{2} + 6 a + 5)
{\arctan (\sqrt{a}) \over \sqrt{a}} - {13 a \over 3} - 5 \biggr] \; .
\end{eqnarray*}

To integrate over $ k $ in Eqs. (\ref{C23}) and (\ref{C21})
we use the Kolmogorov spectrum of the background turbulence,
{\em i.e.,} $ \tau f_{pp}^{(0N)}({\bf k}) = \eta_{T}^{(v)} \varphi(k) ,$
$ \tau h_{pp}^{(0N)}({\bf k}) = \eta_{T}^{(h)} \varphi(k) $ and
$ \mu_{ij}^{(a)}({\bf k}) = \mu_{ij}^{(a)}({\bf R}) \varphi(k) / 3 ,$
where $ \varphi(k) = (\pi k^{2} k_{0})^{-1} (k / k_{0})^{-7/3} ,$
$ \tau(k) = 2 \tau_{0} (k / k_{0})^{-2/3} ,$
where $ k_{0} \leq k \leq k_{d} ,$ $ \quad k_{0} = l_{0}^{-1} ,$
$ l_{0} $ is the maximum scale of turbulent motions and $ k_{d} = k_{0}
{\rm Re}^{3/4} $ is determined by the Kolmogorov's viscous scale
of turbulence.
The integration in $ {\bf k} $-space in Eqs. (\ref{C23}) and
(\ref{C21})  yields
\begin{eqnarray}
\lambda^{(a)}_{ij}(\beta) &=& \Lambda^{(a)}_{ij}(\beta) + \hat \beta_{i}
[\gamma^{(a)}(\beta) \hat \beta_{j}
\nonumber \\
&& + \Psi_{2}(\beta) \mu_{jn}^{(a)} \hat \beta_{n}] \;,
\label{C25} \\
\zeta^{(a)}_{ijmn}(\beta) &=& \xi^{(a)}_{ijmn}(\beta) + \hat \beta_{n}
[U^{(a)}_{ijm}(\beta)
\nonumber \\
&& + \Gamma^{(a)}(\beta) \delta_{ij} \hat \beta_{m}] \;,
\label{D10}
\end{eqnarray}
where
\begin{eqnarray}
\Lambda^{(a)}_{ij}(\beta) &=& \Psi_{1}(\beta) \mu_{ij}^{(a)}
+ \Psi_{2}(\beta) \mu_{in}^{(a)} \beta_{nj} + \delta_{ij} \{ [A_{1}(\beta)
\nonumber \\
& & + {1\over 2} A_{2}(\beta)] \eta_{T}^{(a)} + {1\over 4}
\Psi_{3}(\beta) \mu_{\beta}^{(a)} \} \;,
\label{P43} \\
\gamma^{(a)}(\beta)  &=& {5 \over 12} C_{2}(\beta)
\mu_{\beta}^{(a)} - {1\over 2} A_{2}(\beta) \eta_{T}^{(a)} \;,
\label{P44} \\
\Gamma^{(a)}(\beta)  &=& {5 \over 12} C_{2}(\beta)
\mu_{\beta}^{(a)} + {1\over 2} A_{2}(\beta) \eta_{T}^{(a)} \;,
\label{D11} \\
U^{(a)}_{ijm}(\beta) &=& {5 \over 6}( \{ [A_{2}(\beta) -
C_{3}(\beta)] \mu_{ij}^{(a)} \hat \beta_{m}
\nonumber \\
&& - C_{3}(\beta) (\mu_{im}^{(a)} \hat \beta_{j} + \mu_{ip}^{(a)}
\hat \beta_{p} \delta_{mj}
\nonumber \\
& & - \mu_{mp}^{(a)} \hat \beta_{p} \delta_{ij}) - C_{2}(\beta) \mu_{ip}^{(a)}
\beta_{pjm} \} \;,
\label{D12} \\
\xi^{(a)}_{ijmn}(\beta) &=& {1\over 2} \delta_{mn} \{
[A_{1}(\beta) \eta_{T}^{(a)} + {5 \over 6}  C_{3}(\beta)
\mu_{\beta}^{(a)}] \delta_{ij}
\nonumber \\
&& + {5 \over 3} [A_{1}(\beta) - C_{1}(\beta)] \mu_{ij}^{(a)} - {5
\over 3} C_{3}(\beta) \mu_{ip}^{(a)} \beta_{pj} \}
\nonumber \\
& & + {5 \over 6} \{ \delta_{ij} [C_{1}(\beta) \mu_{mn}^{(a)} +
C_{3}(\beta) \mu_{np}^{(a)} \beta_{pm}]
\nonumber \\
& & - C_{1}(\beta) \mu_{im}^{(a)} \delta_{jn} - C_{3}(\beta) \mu_{ip}^{(a)}
\beta_{pm} \delta_{jn} \}  \;,
\label{D13}
\end{eqnarray}
and $ \mu_{\beta}^{(a)} = \mu_{ps}^{(a)} \beta_{sp} ,$
$ \Psi_{1}(\beta) = (5/6) [A_{1}(\beta) + A_{2}(\beta) + C_{1}(\beta)] ,$
$ \Psi_{2}(\beta) = (5/6) [C_{3}(\beta) - A_{2}(\beta)] ,$ $ \Psi_{3}(\beta)
= (5/3) [A_{2}(\beta) + C_{3}(\beta)] .$ The functions $ A_{n}(\beta)
= \int_{k_{0}}^{\infty} \bar A_{n}(a) \varphi(k) k^{2}
\,d k = (3 \beta^{4} / \pi) \int_{\beta}^{\infty} (\bar A_{n}(X^{2}) / X^{5})
\,d X $ and similarly for $ C_{n}(\beta) ,$ where $ a = [\beta u_{0}
k \tau(k) / 2]^{2} = X^{2} = \beta^{2} (k / k_{0})^{2/3} ,$
and we took into account that the inertial range
of the turbulence exists in the scales: $ l_{d} \leq r \leq l_{0} .$ Here
the maximum scale of the turbulence $ l_{0} \ll L_{B} ,$ and $ l_{d} = l_{0} /
{\rm Re}^{3/4} $ is the viscous scale of turbulence, and
$ L_{B} $ is the characteristic scale of variations of the nonuniform
mean magnetic field. For very large Reynolds
numbers $ k_{d} = l_{d}^{-1} $ is very large and the turbulent
hydrodynamic and magnetic energies are very small in the viscous
dissipative range of the turbulence $ 0 \leq r \leq l_{d} .$
Thus we integrated in $ A_{n} $ over $ k $ from $ k_{0} = l_{0}^{-1} $ to
$ \infty .$
The functions $ A_{n}(\beta) $ and $ C_{n}(\beta) $ are given in Appendix B.
In Eqs. (\ref{D10})--(\ref{D13}) we omitted terms which are
symmetric in indexes $ i $ and $ n $ because after multiplication
$ \zeta^{(a)}_{ijmn}(\beta) $ by $ \varepsilon_{lin} $ these
symmetric terms vanish [see Eq. (\ref{C20})].

In order to extract terms $ \propto \varepsilon_{ijm} \hat \beta_{m} $
which contribute to the nonlinear diamagnetic and paramagnetic velocities,
we split $ b_{ijk} $ into two parts, {\em i.e.,} $ b_{ijk} = b_{ijk}^{(1)}
+ b_{ijk}^{(2)} ,$ where
\begin{eqnarray}
b_{ijk}^{(1)} &=& \varepsilon_{inm} \hat \beta_{m} \{ \delta_{jn}
[\gamma^{(P)}(\beta) \hat \beta_{k} + \Psi_{2}(\beta) \mu_{kp}^{(P)}
\hat \beta_{p}]
\nonumber \\
& & + 2 [\Gamma^{(C)}(\beta) \delta_{nk} \hat \beta_{j}
+ U_{nkj}^{(C)}(\beta)] \} \;,
\label{D14} \\
b_{ijk}^{(2)} &=& \varepsilon_{ijn} \Lambda^{(P)}_{nk}(\beta) +
2 \varepsilon_{inm} \xi^{(C)}_{nkjm}(\beta) \;
\label{D15}
\end{eqnarray}
[see the definitions given by Eqs. (\ref{DD1})-(\ref{DD3})].
Next, we calculate $ b_{ijk} B_{j,k} .$ Using Eqs. (\ref{P20}), (\ref{D14})
and (\ref{D15}) we also split the electromotive force into two parts
\begin{eqnarray}
\bec{\cal E} &=& \bec{\cal E}^{(1)} + \bec{\cal E}^{(2)} \;,
\label{DD15} \\
{\cal E}^{(1)}_{i} &=& b_{ijk}^{(1)} B_{j,k} \;,
\label{D15D} \\
{\cal E}^{(2)}_{i} &=& a_{ij} B_{j} + b_{ijk}^{(2)} B_{j,k} \; .
\label{D14D}
\end{eqnarray}
Using Eqs. (\ref{D14}) and (\ref{D15D}) we obtain
\begin{eqnarray}
{\cal E}^{(1)}_{i} = ({\bf V}^{(N)} {\bf \times} {\bf B})_{i}
- \eta^{(1)}_{ij} (\bec{\bf \nabla} {\bf \times} {\bf B})_{j} \;,
\label{D18}
\end{eqnarray}
where
\begin{eqnarray}
&& V^{(N)}_{i}({\bf B}) = {1 \over 2 B^{2}} [\gamma^{(P)}(\beta) +
2 \Gamma^{(C)}(\beta)] \nabla_{i} B^{2}
\nonumber \\
& & + {1 \over B} [2 U^{(C)}_{ikj}(\beta) + \Psi_{2}(\beta)
\mu_{kp}^{(P)} \hat \beta_{p} \delta_{ij}] \nabla_{k} B_{j} \;,
\label{D19} \\
&&\eta^{(1)}_{ij} = \gamma^{(P)}(\beta) P_{ij}(\beta) \;,
\label{D19D}
\end{eqnarray}
and $ P_{ij}(\beta) = \delta_{ij} - \beta_{ij} .$
For the calculation of the terms $ \propto \gamma^{(P)}(\beta) $ in these
equations we used an identity
$ \varepsilon_{imn} \beta_{np} B_{m,p} \equiv - [{\bf B} {\bf \times}
({\bf B} {\bf \cdot} \bec{\nabla}) {\bf B}]_{i} / B^{2}
= - [{\bf B} {\bf \times} \bec{\nabla} (B^{2} / 2)]_{i} / B^{2}
- P_{ip}(\beta) (\bec{\nabla} {\bf \times} {\bf B})_{p} ,$
which follows from the formula $ ({\bf B} {\bf \cdot} \bec{\nabla}) {\bf B}
= (1/2) \bec{\nabla} B^{2} - {\bf B} {\bf \times} (\bec{\nabla} {\bf \times}
{\bf B}) .$
Following to \cite{R80} we use an identity $ B_{j,k} = (\partial \hat B)_{jk}
- \varepsilon_{jkl} (\bec{\bf \nabla} {\bf \times} {\bf B})_{l} / 2 $
in order to rewrite Eq. (\ref{D14D}) in the form
\begin{eqnarray}
{\cal E}^{(2)}_{i} &=& {\alpha}_{ij} B_{j} + ({\bf U} {\bf \times}
{\bf B})_{i} - {\eta}^{(2)}_{ij} (\bec{\nabla} {\bf \times} {\bf
B})_{j}
\nonumber \\
& & - {\kappa}_{ijk} ({\partial \hat B})_{jk} \;, \label{P23}
\end{eqnarray}
where
\begin{eqnarray}
\eta_{ij}^{(2)} &=& (\varepsilon_{ikp} b_{jkp}^{(2)} +
\varepsilon_{jkp} b_{ikp}^{(2)}) / 4 \;,
\label{DD40} \\
\quad \kappa_{ijk}({\bf B}) &=& - (b_{ijk}^{(2)} + b_{ikj}^{(2)}) / 2 \; .
\label{D40}
\end{eqnarray}

Using Eqs. (\ref{D18}) and (\ref{P23}) we obtain the equation for the
electromotive force $ \bec{\cal E} = \bec{\cal E}^{(1)} + \bec{\cal
E}^{(2)} $ which is given by Eq. (\ref{D27}).
The tensor of turbulent magnetic diffusion,
\begin{eqnarray}
\eta_{ij}({\bf B}) = \eta_{ij}^{(1)} + \eta_{ij}^{(2)} \;,
\label{DD41}
\end{eqnarray}
is given by
\begin{eqnarray}
\eta_{ij}({\bf B}) &=& \delta_{ij} \{ A_{1}(\beta) \eta_{T}^{(P)}
+ {5 \over 12} [C_{2}(\beta) + 2 C_{3}(\beta)] \mu_{\beta}^{(P)}
\nonumber \\
& & - [A_{1} \eta_{T}]^{(C)} - {5 \over 6} [C_{3}
\mu_{\beta}]^{(C)} \} - {1 \over 4} [2 \Psi_{1}(\beta)
\mu_{ij}^{(P)}
\nonumber \\
& & + \Psi_{2}(\beta) \bar \mu_{ij}^{(P)}] + {5 \over 6} [(A_{1} +
C_{1}) \mu_{ij}]^{(C)}
\nonumber \\
& & + {5 \over 12}  [C_{3} \bar \mu_{ij}]^{(C)} + {1 \over 2}
\beta_{ij} [A_{2}(\beta) \eta_{T}^{(P)}
\nonumber \\
& & -  {5 \over 6} C_{2}(\beta) \mu_{\beta}^{(P)}] \;, \label{D20}
\end{eqnarray}
where $ \bar \mu_{ij}^{(a)} = \mu_{in}^{(a)} \beta_{nj} + \beta_{in}
\mu_{nj}^{(a)} ,$ and we used Eqs. (\ref{D19D}), (\ref{DD40}) and the definitions
(\ref{DD1})-(\ref{DD3}). In particular,
$ [X]^{(C)}(\beta) = X^{(v)}(\beta) - X^{(v)}(\sqrt{2}\beta) +
X^{(h)}(\sqrt{2}\beta) $ which implies, {\em e.g.,}
$ [A_{1} \eta_{T}]^{(C)} = A_{1}(\beta) \eta_{T}^{(v)} - A_{1}(\sqrt{2}\beta)
\eta_{T}^{(v)} + A_{1}(\sqrt{2}\beta) \eta_{T}^{(h)} .$

Using Eqs. (\ref{D15}) and (\ref{D40}) we calculate $ \kappa_{ijk}({\bf B}) :$
\begin{eqnarray}
&&\kappa_{ijk}({\bf B}) = - {1 \over 2} [\Psi_{1}(\beta) \hat
L_{ijk}^{(P)} + \Psi_{2}(\beta) \hat N_{ijk}^{(P)}]
\nonumber \\
& & + {5 \over 6}  [(A_{1} - 3 C_{1}) \hat L_{ijk}]^{(C)} - {5
\over 2} [C_{3} \hat N_{ijk}]^{(C)} \;, \label{D24}
\end{eqnarray}
where
$ \hat L_{ijk}^{(a)} = \varepsilon_{ijn} \mu_{nk}^{(a)} +
\varepsilon_{ikn} \mu_{nj}^{(a)} ,$ $ \quad \hat N_{ijk}^{(a)} =
\mu_{np}^{(a)} (\varepsilon_{ijn} \beta_{pk} +
\varepsilon_{ikn} \beta_{pj}) $ and $ [C_{3} \hat N_{ijk}]^{(C)}
= C_{3}(\beta) \hat N_{ijk}^{(v)} - C_{3}(\sqrt{2}\beta) (\hat N_{ijk}^{(v)}
- \hat N_{ijk}^{(h)}) $ and similarly for $ [(A_{1} - 3 C_{1})
\hat L_{ijk}]^{(C)} $ [see Eq. (\ref{DD1})].

The asymptotic formulas for the nonlinear coefficients defining the
mean electromotive force for $ \beta \ll 1 $ are given by
\begin{eqnarray}
\eta_{ij}({\bf B}) &=& \delta_{ij} \eta_{T}^{(v)} - {1 \over 2} \mu_{ij}^{(M)}
- {2 \over 5} \beta^{2} \biggl[ \delta_{ij} \biggl(2 \eta_{T}^{(v)} -
\eta_{T}^{(h)}
\nonumber \\
& & + {5 \over 21} (2 \mu_{\beta}^{(v)} - \mu_{\beta}^{(h)})
\biggr) + {5 \over 42} (5 \mu_{ij}^{(h)} - 19 \mu_{ij}^{(v)}
\nonumber \\
& & + 2 \bar \mu_{ij}^{(v)} + 5 \bar \mu_{ij}^{(h)}) + \beta_{ij}
\eta_{T}^{(P)} \biggr] \;,
\label{D21} \\
\alpha_{ij}^{(v)}({\bf B}) &=& \alpha_{0}^{(v)} \delta_{ij} +
\nu_{ij} - (2/5) \beta^{2} \{\delta_{ij} [3 \alpha_{0}^{(v)}
+ \nu_{\beta} (1
\nonumber \\
& & + (8/7) \epsilon)] + \nu_{ij} (2 - (9/7) \epsilon)\} \;,
\label{C5} \\
\alpha_{ij}^{(h)}({\bf B}) &=& \alpha_{0}^{(h)}({\bf B}) (1 - 3 \beta^{2} / 5)
\delta_{ij}  \;,
\label{D30}
\end{eqnarray}
and for $ \beta \gg 1 $ they are given by:
\begin{eqnarray}
\eta_{ij}({\bf B}) &=& {3 \pi \over 5 \beta} \biggl [ \delta_{ij}
\biggl( {\eta_{T}^{(M)} \over \sqrt{2}} + \eta_{T}^{(h)} + {5
\over 48} [\mu_{\beta}^{(v)} (3 - \sqrt{2})
\nonumber \\
& & + \mu_{\beta}^{(h)} (\sqrt{2} + 1)] \biggr) + {5 \over 48}
\biggl( \mu_{ij}^{(v)}(9 - 5 \sqrt{2})
\nonumber \\
& & + \mu_{ij}^{(h)}(5 \sqrt{2} - 1 ) - {3 \over 10} [\bar
\mu_{ij}^{(v)} (5 - \sqrt{2})
\nonumber \\
& & + \bar \mu_{ij}^{(h)}(3 + \sqrt{2})] \biggr) - {1 \over 2}
\beta_{ij} \biggl( \eta_{T}^{(P)}
\nonumber \\
& & + {5 \over 8} \mu_{\beta}^{(P)} \biggr) \biggr ] \;,
\label{D22} \\
\alpha_{ij}^{(v)}({\bf B}) &=& - {3 \pi \over 10 \beta}
[\delta_{ij} (1 - \epsilon) \nu_{\beta} - \nu_{ij} (1 + 9
\epsilon)]
\nonumber \\
& & + {2 \alpha_{0}^{(v)} \over \beta^{2}} \delta_{ij} \;,
\label{C6} \\
\alpha_{ij}^{(h)}({\bf B}) &=& {3 \pi \over 2 \beta^{2}}
\alpha_{0}^{(h)}({\bf B}) \delta_{ij} \; . \label{D32}
\end{eqnarray}
The asymptotic formulas for the tensor $ \kappa_{ijk} $  for $ \beta \ll 1 $
and $ \beta \gg 1 $ are given by Eqs. (\ref{PD25}) and (\ref{PD30}).

\section{The functions $ A_{\alpha}(\beta) $ and $ C_{\alpha}(\beta) $}

The functions $ A_{\alpha}(\beta) $ and $ C_{\alpha}(\beta) $ are given by
\begin{eqnarray*}
A_{1}(\beta) &=& {6 \over 5} \biggl[{\arctan \beta
\over \beta} \biggl(1 + {5 \over 7 \beta^{2}} \biggr) + {1 \over 14} L(\beta)
- {5 \over 7\beta^{2}} \biggr]  \;,
\\
A_{2}(\beta) &=& - {6 \over 5} \biggl[{\arctan \beta
\over \beta} \biggl(1 + {15 \over 7 \beta^{2}} \biggr) - {2 \over 7} L(\beta)
- {15 \over 7\beta^{2}} \biggr]  \;,
\\
C_{1}(\beta) &=& {3 \over 10} \biggl[{\arctan \beta \over \beta}
\biggl(1 + {10 \over 7 \beta^{2}} + {5 \over 9 \beta^{4}}
\biggr) + {2 \over 63} L(\beta)
\nonumber \\
& & - {235 \over 189 \beta^{2}} - {5 \over 9 \beta^{4}} \biggr]
\;,
\\
C_{2}(\beta) &=& {3 \over 2} \biggl[{\arctan \beta \over \beta}
\biggl({3 \over 5} + {30 \over 7 \beta^{2}} + {35 \over 9 \beta^{4}}
\biggr) + {16 \over 315} L(\beta)
\nonumber \\
& & - {565 \over 189 \beta^{2}} - {35 \over 9 \beta^{4}} \biggr]
\;,
\\
C_{3}(\beta) &=& - {3 \over 2} \biggl[{\arctan \beta \over \beta}
\biggl({1 \over 5} + {6 \over 7 \beta^{2}} + {5 \over 9 \beta^{4}}
\biggr) - {8 \over 315} L(\beta)
\nonumber \\
& & - {127 \over 189 \beta^{2}} - {5 \over 9 \beta^{4}} \biggr]
\;,
\end{eqnarray*}
where $ L(\beta) = 1 - 2 \beta^{2} + 2 \beta^{4} \ln (1 + \beta^{-2}) .$
For $ \beta \ll 1 $ these functions are given by
\begin{eqnarray*}
A_{1}(\beta) &\sim& 1 - {2 \over 5} \beta^{2}  \;, \quad
A_{2}(\beta) \sim - {4 \over 5} \beta^{2} \;,
\\
C_{1}(\beta) &\sim& {1 \over 5}  [1 - {2 \over 7} \beta^{2}]  \;,
\quad C_{2}(\beta) \sim - {32 \over 105} \beta^{4} \ln \beta \;,
\\
C_{3}(\beta) &\sim& - {4 \over 35} \beta^{2} \;,
\end{eqnarray*}
and for $ \beta \gg 1 $ they are given by
\begin{eqnarray*}
A_{1}(\beta) &\sim& {3 \pi \over 5 \beta} - {2 \over \beta^{2}}
\;, \quad A_{2}(\beta) \sim - {3 \pi \over 5 \beta} + {4 \over
\beta^{2}} \;,
\\
C_{1}(\beta) &\sim& {3 \pi \over 20 \beta} \;, \quad C_{2}(\beta)
\sim {9 \pi \over 20 \beta} \;,
\\
C_{3}(\beta) &\sim& - {3 \pi \over 20 \beta} \; .
\end{eqnarray*}
Since the function $ A_{1}(\beta) + A_{2}(\beta) \sim O(\beta^{-2}) $
for $ \beta \gg 1 $
(it describes an isotropic part of the $ \alpha $-effect) we took into
account in the functions $ A_{1}(\beta) $ and $ A_{2}(\beta) $ the terms
which are of the order of $ \sim O(\beta^{-2}) .$ Here we also used
that for $ \beta \ll 1 $ the function $ L(\beta) \sim 1 - 2 \beta^{2}
- 4 \beta^{4} \ln \beta ,$ and for $ \beta \gg 1 $ the function
$ L(\beta) \sim 2 / 3 \beta^{2} .$

\section{Derivation of the nonlinear dependencies $ \eta_{A}(B) ,$
$ \eta_{B}(B) $ and $ V_{A}(B) $}

Now we consider an anisotropic background turbulence with one preferential
direction, say along unit vector $ {\bf e} ,$ where
$ {\bf e} {\bf \cdot} \bec{\hat \beta} = 0 .$
In this case
\begin{eqnarray}
{\bf V}^{(N)}  &=& B^{-2} [V^{(1)} \bec{\nabla} B^{2} + V^{(2)}
{\bf e} ({\bf e} {\bf \cdot} \bec{\nabla}) B^{2}
\nonumber \\
& & + V^{(3)} ({\bf B} {\bf \cdot} \bec{\nabla}) {\bf B}] \;,
\label{DD48} \\
{\bf U}  &=& B^{-2} [U^{(1)} \bec{\nabla} B^{2} + U^{(2)} {\bf e}
({\bf e} {\bf \cdot} \bec{\nabla}) B^{2}
\nonumber \\
& & + U^{(3)} ({\bf B} {\bf \cdot} \bec{\nabla}) {\bf B}] \;,
\label{DD49} \\
\kappa_{ijk} ({\partial \hat B})_{jk} &=& - B^{-2} \{ \tilde W
[\bec{\nabla} B^{2} + 2 ({\bf B} {\bf \cdot} \bec{\nabla}) {\bf
B}] {\bf \times} {\bf B}
\nonumber \\
& & - M_{\kappa} {\bf e} {\bf \times} ({\bf e} {\bf \cdot}
\bec{\nabla}) {\bf B} \}_{i} \;, \label{DD50}
\end{eqnarray}
where
\begin{eqnarray*}
V^{(1)} &=& - {1 \over 4} \{ A_{2}(\beta) \eta_{T}^{(P)} + {5
\over 18} C_{2}(\beta) \varepsilon_{\mu}^{(P)}
\\
& & + {5 \over 9} [(C_{2} + 2 A_{2}) \varepsilon_{\mu}]^{(C)} - {1
\over 2}  [A_{2} \eta_{T}]^{(C)} \} \;,
\\
V^{(2)} &=& {5 \over 6}  [(A_{2} - C_{3}) \varepsilon_{\mu}]^{(C)}
\;,
\\
V^{(3)} &=& {5 \over 9} [C_{3} \varepsilon_{\mu}]^{(C)} - {1 \over
3} \Psi_{2}(\beta) \varepsilon_{\mu}^{(P)} \;,
\\
U^{(1)} &=& - (\sqrt{2} \beta / 48) \Psi(\sqrt{2} \beta) \;,
\\
U^{(2)} &=& - (\sqrt{2} \beta / 4) \Psi'_{1}(\sqrt{2} \beta)
\varepsilon_{\mu}^{(M)} \;,
\\
U^{(3)} &=& {1 \over 6} \Psi_{2}(\sqrt{2} \beta)
\varepsilon_{\mu}^{(M)}  \;,
\\
\tilde W &=& - {1 \over 12} \{ \Psi_{2}(\beta)
\varepsilon_{\mu}^{(P)} + 5 [C_{3} \varepsilon_{\mu}]^{(C)} \} \;
.
\end{eqnarray*}
In order to derive Eq. (\ref{DD50}) we used the following identities:
\begin{eqnarray*}
L_{ijk}^{(a)} ({\partial \hat B})_{jk} &=& - \varepsilon_{\mu}^{(a)}
[{\bf e} {\bf \times} ({\bf e} {\bf \cdot} \bec{\nabla}) {\bf B}]_{i} \;,
\\
N_{ijk}^{(a)} ({\partial \hat B})_{jk} &=& - {1 \over 6 B^{2}}
\varepsilon_{\mu}^{(a)} [(\bec{\nabla} B^{2} + 2 ({\bf B} {\bf
\cdot} \bec{\nabla}) {\bf B}) {\bf \times} {\bf B}]_{i} \; .
\end{eqnarray*}
Using Eqs. (\ref{DD48})-(\ref{DD50}) and (\ref{D20}) we calculate the functions
$ M_{\eta} ,$ $ M_{e} ,$ $ M_{\beta} ,$ $ M_{\kappa} ,$ $ M_{V}^{(1)} $ and
$ M_{V}^{(2)} $ in Eqs. (\ref{D47})-(\ref{D49}):
\begin{eqnarray}
M_{\eta} &=& A_{1}(\beta) \eta_{T}^{(P)} + {5\over 36}
[A_{1}(\beta) + 4 A_{2}(\beta) + C_{1}(\beta)
\nonumber \\
& & - C_{2}(\beta) - 5 C_{3}(\beta)] \varepsilon_{\mu}^{(P)} -
{5\over 18}  [(C_{1} + A_{1}) \varepsilon_{\mu}]^{(C)}
\nonumber \\
& & - [A_{1} \eta_{T}]^{(C)} + {1 \over 6} \Psi_{2}(\sqrt{2}
\beta) \varepsilon_{\mu}^{(M)} \;,
\label{D50} \\
M_{\kappa} &=& {1 \over 2} \Psi_{1}(\beta) \varepsilon_{\mu}^{(P)}
+ {5\over 6} [(3 C_{1} - A_{1}) \varepsilon_{\mu}]^{(C)} \;,
\label{D51} \\
M_{\beta} &=& {1 \over 6} \{3 A_{2}(\beta) \eta_{T}^{(P)} +
[{5\over 6} C_{2}(\beta) + 4 \Psi_{2}(\beta)]
\varepsilon_{\mu}^{(P)}
\nonumber \\
& & - \Psi_{2}(\sqrt{2} \beta) \varepsilon_{\mu}^{(M)} \} \;,
\label{D52} \\
M_{e} &=& - {1 \over 2} \Psi_{1}(\beta) \varepsilon_{\mu}^{(P)} +
{5\over 6} [(C_{1} + A_{1}) \varepsilon_{\mu}]^{(C)}  \;,
\label{DD51} \\
M_{V}^{(1)} & \equiv & V^{(1)} + U^{(1)} + 2 W + {1 \over 2}
(V^{(3)} + U^{(3)})
\nonumber \\
&=& - {1 \over 4} A_{2}(\beta) \eta_{T}^{(P)} - {5\over 72}
[C_{2}(\beta) + 4 C_{3}(\beta)
\nonumber \\
& & - 4 A_{2}(\beta)] \varepsilon_{\mu}^{(P)} + {1 \over 12}
\Psi_{2}(\sqrt{2} \beta) \varepsilon_{\mu}^{(M)}
\nonumber \\
& & + {5\over 36} \{ [A_{2} \eta_{T}]^{(C)} - [(C_{2} + 2 A_{2} +
4 C_{3}) \varepsilon_{\mu}]^{(C)} \}
\nonumber \\
& & - (\sqrt{2} \beta / 48) \Psi(\sqrt{2} \beta) \;,
\label{D53} \\
M_{V}^{(2)} & \equiv & V^{(2)} + U^{(2)} = {5\over 6} [(A_{2} -
C_{3}) \varepsilon_{\mu}]^{(C)}
\nonumber \\
& & - {\sqrt{2} \beta \over 4} \Psi'_{1}(\sqrt{2} \beta)
\varepsilon_{\mu}^{(M)} \; . \label{D54}
\end{eqnarray}
Now we take into account that $ {\bf V}^{(N)} ,$ $ {\bf U} $ and
$ \bec{\kappa} $ contribute into the tensor $ \eta_{ij} .$
This implies that in order to calculate $ M_{\eta} ,$ $ M_{e} ,$
and $ M_{\beta} $ we perform the change
\begin{eqnarray}
\eta_{ij} \to \eta_{ij} + P_{ij}(\beta) [V^{(3)} + U^{(3)} + 2W] \;,
\label{DD54}
\end{eqnarray}
where the second term in (\ref{DD54}) [which is proportional to
$ P_{ij}(\beta) ]$ describes a contribution $ {\bf V}^{(N)} ,$ $ {\bf U} $
and $ \bec{\kappa} $ into the tensor $ \eta_{ij} .$
Using Eqs. (\ref{D47})-(\ref{D49}) and (\ref{D50})-(\ref{D54}) we calculate
the functions $ \eta_{A}(B) ,$ $ \eta_{B}(B) $ and $ V_{A}(B) $:
\begin{eqnarray}
\eta_{A}(B) &=& \tilde \eta(B) + (10/9) [(2 C_{1} - A_{1})
\varepsilon_{\mu}]^{(C)}
\nonumber \\
& & - [A_{1} \eta_{T}]^{(C)} \;,
\label{D61} \\
\eta_{B}(B) &=& \tilde \eta(B) + {5\over 18} [(8 C_{1} + C_{2} +
10 C_{3} - 4 A_{1}
\nonumber \\
& & - 4 A_{2}) \varepsilon_{\mu}]^{(C)} - [(A_{1} + A_{2})
\eta_{T}]^{(C)}
\nonumber \\
& & + {\sqrt{2} \beta \over  24} \Psi(\sqrt{2} \beta) \;,
\label{D62}\\
V_{A}(B) &=& \{{5\over 18} [(4 A_{2} - C_{2} - 10 C_{3})
\varepsilon_{\mu}]^{(C)} + [A_{2} \eta_{T}]^{(C)}
\nonumber \\
& & - {\sqrt{2} \beta \over  24} \Psi(\sqrt{2} \beta) \} (\ln
|B|)' \;, \label{D90}
\end{eqnarray}
where $ \Psi(x) = 12 [A'_{1}(x) + (1/2) A'_{2}(x)] \eta_{T}^{(M)}
+ \Psi'_{0}(x) \varepsilon_{\mu}^{(M)} ,$
$ \quad \tilde \eta(B) = [A_{1}(\beta) + (1/2) A_{2}(\beta)] \eta_{T}^{(P)} +
(\varepsilon_{\mu}^{(P)} / 12) \Psi_{0}(\beta) ,$ $ \quad \Psi_{0}(\beta)
= (5/3) [4 A_{1}(\beta) + 3 A_{2}(\beta) + 4 C_{1}(\beta)
- C_{3}(\beta)] $ and $ [X]^{(C)} $ is defined by Eq. (\ref{DD1}).
The asymptotic formulas for the functions $ \eta_{A} ,$ $ \eta_{B} ,$
$ V_{A} $ and $ \alpha_{ij}^{(v)} $ for $ \beta \ll 1 $ are given by
\begin{eqnarray}
\eta_{A}(B) &=& \eta_{\ast} - {2 \over 5} \beta^{2} [3
\eta_{T}^{(v)} + {10\over 63} (14 \varepsilon_{\mu}^{(v)}
\nonumber \\
& & - \varepsilon_{\mu}^{(h)})] \;,
\label{P61}\\
\eta_{B}(B) &=& \eta_{\ast} - {2 \over 5} \beta^{2} [9
\eta_{T}^{(v)} - 8 \eta_{T}^{(h)} + {10\over 63} (41
\varepsilon_{\mu}^{(v)}
\nonumber \\
& & - 37 \varepsilon_{\mu}^{(h)})]  \;,
\label{P62}\\
V_{A}(B) &=& {4 \over 5} \beta^{2} [3 \eta_{T}^{(v)} - 4
\eta_{T}^{(h)} + {5 \over 7} (3 \varepsilon_{\mu}^{(v)}
\nonumber \\
& & - 4 \varepsilon_{\mu}^{(h)})] (\ln |B|)'  \;,
\label{P90}\\
\alpha_{ij}^{(v)}({\bf B}) &=& \delta_{ij} [(\alpha_{0}^{(v)} -
(1/3) \varepsilon_{\alpha}) (1 - {6 \over 5} \beta^{2})
\nonumber \\
& & - {2\over 105} \beta^{2} \varepsilon_{\alpha} \epsilon] \;,
\label{C9}
\end{eqnarray}
and for $ \beta \gg 1 $ they are given by
\begin{eqnarray}
\eta_{A}(B) &=& {\pi \over 6 \beta} [{9 \over 5} (\sqrt{2} - 1)
\eta_{T}^{(M)} + (\sqrt{2} - {7\over 8}) \varepsilon_{\mu}^{(v)}
\nonumber \\
& & - (\sqrt{2} - {9\over 8}) \varepsilon_{\mu}^{(h)}] \;,
\label{R61} \\
\eta_{B}(B) &=& {\pi \over 4 \sqrt{2} \beta} \{ {3 \over 5} [(2
\sqrt{2} - 1) \eta_{T}^{(v)} + (2 \sqrt{2} + 1) \eta_{T}^{(h)}]
\nonumber \\
& & + {1\over 24} [(22 \sqrt{2} - 13) \varepsilon_{\mu}^{(v)} +
(18 \sqrt{2}
\nonumber \\
& & + 13) \varepsilon_{\mu}^{(h)}] \} \;,
\label{R62}\\
V_{A}(B) &=& - {3 \pi \over 4 \sqrt{2} \beta} \biggl[ \biggl({4
\sqrt{2} \over 5} - 1 \biggr) (\eta_{T}^{(v)} + {5\over 8}
\varepsilon_{\mu}^{(v)}) + \eta_{T}^{(h)}
\nonumber \\
& & + {5\over 8} \varepsilon_{\mu}^{(h)} \biggr] (\ln |B|)'  \;,
\label{R90}\\
\alpha_{ij}^{(v)}({\bf B}) &=& - \delta_{ij} \{ {\pi \over \beta}
\varepsilon_{\alpha} \epsilon - {2 \over \beta^{2}}
[\alpha_{0}^{(v)} - {1\over 3} \varepsilon_{\alpha} (1 -
\epsilon)] \} \; . \label{C10}
\end{eqnarray}

\end{document}